\documentclass[prb,aps,twocolumn,showpacs,amsmath,amssymb,floatfix]{revtex4}
\usepackage{subfigure}
\usepackage{graphicx}
\usepackage{dcolumn}
\usepackage{bm}
\usepackage{wasysym}
\usepackage{here}
\usepackage{textcomp} 
\def\beq{\begin{equation}}
\def\eeq{\end{equation}}
\def\bea{\begin{eqnarray}}
\def\eea{\end{eqnarray}}
\def\nn{\nonumber}

\begin{document}
\preprint{APS/123-QED}
\title{Dynamical obstruction in a constrained system and its realization
  in lattices of superconducting devices}
\author{Claudio Castelnovo$^1$, Pierre Pujol$^{1,2}$, and Claudio Chamon$^1$}
\affiliation{
$^1$ Physics Department, Boston University, Boston, MA 02215, USA
\\
$^2$ Laboratoire de Physique, Groupe de Physique Th{\'e}orique de
l'{\'E}cole Normale Sup{\'e}rieure, Lyon, France}
\date{\today}

\begin{abstract}
Hard constraints imposed in statistical mechanics models can lead to
interesting thermodynamical behaviors, but may at the same time raise
obstructions in the thoroughfare to thermal equilibration.  Here we
study a variant of Baxter's 3-color model in which local interactions
and defects are included, and discuss its connection to triangular
arrays of Josephson junctions of superconductors with broken
time-reversal symmetry and \textit{kagom\'e} networks of
superconducting wires. The model is equivalent to an Ising model in a
hexagonal lattice with the additional constraint that the
magnetization of each hexagon is $\pm 6$ or $0$. Defects in the
superconducting models correspond to violations of this constraint,
and include fractional and integer vortices, as well as open strings
within 2-color loops. In the absence of defects, and for ferromagnetic
interactions, we find that the system is critical for a range of
temperatures (critical line) that terminates when it undergoes an
exotic first order phase transition with a jump from a zero
magnetization state into the fully magnetized state at finite
temperature.  Dynamically, however, we find that the system becomes
frozen into domains.  The domain walls are made of perfectly straight
segments, and domain growth appears frozen within the time scales
studied with Monte Carlo simulations, with the system trapped into a
``polycrystalline'' phase. This dynamical obstruction has its origin
in the topology of the allowed reconfigurations in phase space, which
consist of updates of closed loops of spins. Only an extreme rare
event dominated proliferation of confined defects may overcome this
obstruction, at much longer time scales. Also as a consequence of the
dynamical obstruction, there exists a dynamical temperature, lower
than the (avoided) static critical temperature, at which the system is
seen to jump from a ``supercooled liquid'' to the ``polycrystalline''
phase within our Monte Carlo time scale. In contrast, for
antiferromagnetic interactions, we argue that the system orders for
infinitesimal coupling because of the constraint, and we observe no
interesting dynamical effects.
\end{abstract}
\pacs{00000}
\maketitle
%
%
\section{Introduction}
Systems with hard constraints often 
display interesting thermodynamic properties such as infinite order phase 
transitions or, on the contrary, very sharp first order phase 
transitions. Many of these models can be described in terms of vertex 
models and some of them are exactly solvable. Examples of such systems are 
given by dimer models~\cite{Kasteleyn}, the planar ice model~\cite{Lieb} or 
the three coloring model of the hexagonal lattice~\cite{Baxter}. 

It is very natural to ask whether the hard constraint, which leads to
the interesting thermodynamics, may at the same time pose obstructions
in the (possible?) path to thermal equilibration. In essence,
equilibrium properties require averages over all the configurations
allowed by the constraint, weighted in accordance with the appropriate
Boltzmann-Gibbs distribution. Dynamically, the system must sample the
different allowed states in a manner that satisfies detailed
balance. However, leaping from an allowed configuration to another
might require large rearrangements, and physically one must
investigate which mechanisms could possibly lead to these moves in
phase space and what are the corresponding time scales.  Sometimes
the constraint forbids any local rearrangement of the system (as in
the present case), and it ought to be softened in order to allow for a
local dynamics. The system then evolves by formation of
constraint-violating defects that propagate and recombine.
 
Plenty of issues arise regarding the dynamical generation and 
recombination of defects, which depend on the microscopic details of 
the physical system, and the energetics of the states outside the 
manifold of constraint-satisfying states. For example, paying the 
energy cost to create a defect already slows down the dynamics; 
however, this waiting for the defect generation simply rescales the 
time scales for dynamical evolution in a trivial way. More interesting 
are those issues related to the possible energy costs for moving 
defects around. In particular, if the microscopics are such that the 
defects (when created in pairs) are confined, one would expect further 
and non-trivial slowing down of the dynamics. 
 
Glassy behavior in constrained 3-color models with infinite range 
interactions has indeed been recently found by Chakraborty, Das, and 
Kondev~\cite{Chakraborty}. This is an interesting example of glassy 
behavior in a Hamiltonian model without quenched disorder, where it was 
found that the characteristic time scales obeyed a Vogel-Fulcher law as 
the temperature approached a dynamical transition temperature, mimicking 
fragile structural glasses. In order to maneuver within the phase 
space of allowed states, non-local loop dynamics was implemented. 
 
In this paper, we study variations of the Baxter 3-color model with 
short range interactions and discuss the possible mechanism for defect 
motion. In particular, we argue that the loop updates used by Chakraborty 
\textit{et al.}~\cite{Chakraborty} correspond to 
the unbinding of certain defect pairs that are deconfined, and thus
they are the least costly mechanism for dynamical evolution. We find
that finite range ferromagnetic interactions lead to a frozen
``polycrystal'', as opposed to a fragile glass as in the case of
infinite range interactions. We present two possible experimental
realizations using lattice arrays of superconducting devices that
could in principle be experimental settings for studying sluggish
relaxation or non-equilibrium effects in Hamiltonian systems without
quenched disorder.
 
In section \ref{sec:model} we present in detail the 3-color model, and
show that it is equivalent to an Ising model on a hexagonal lattice,
with the constraint that the magnetization of each hexagon must be
$\pm 6$ or $0$. In the Ising language the extra interaction that we
add to the 3-color model has a simple form: it is a nearest neighbor
spin-spin interaction. Such interaction is present in the possible
experimental realizations of the model in two different 2-D
superconducting geometries. Because of the constraint imposed on the
plaquettes, the system is critical in the absence of two-spin
interactions ($J=0$) and is described by a $c=2$ Conformal Field
Theory (CFT)~\cite{Kondev}. In Sec.~\ref{sec:thermodyn}, we use this
description to argue about the behavior of the model in the presence
of non-zero two-spin interactions. While for arbitrarily small
antiferromagnetic coupling ($J<0$) the system orders, it remains
critical for small ferromagnetic coupling ($J>0$). The CFT description
near the $J=0$ point is ill-suited for strong couplings. In this
regime we use instead a Cluster Mean Field Method (CMFM) which has
proven to be very accurate in describing constrained system such as 
the ice model~\cite{Vaks}. 
We find a strong first order phase transition where the system 
jumps from the disordered configuration to the Fully Magnetized 
Ferromagnetic State (FMFS). 
 
When the hard constraint is softened, defects are allowed in the
system at a high energy scale $U$, which enters in the defect
formation energy and in the defect pair interactions.  In
Sec.~\ref{sec:defects}, we discuss the role of these defects and their
implications in the dynamics of the system. In the superconducting
realizations there are a number of different defects: fractional
vortices, integer vortices, and open segments of closed two-color
loops. Integer and fractional vortices can be shown to be confined
below a Kosterlitz-Thouless transition temperature that can be rather
high depending on the energy scale $U$.  
Thus, these defects are rather ineffective as a mechanism
to move from one allowed state to another. We show, on the other hand,
that the end points of open segments of closed loops made of two
alternating colors are deconfined, they can move around and travel a
whole closed loop, and therefore they are the main actors for the
evolution of the system.  For defect formation rates much smaller than
the defect recombination rates, this evolution corresponds essentially
to the loop dynamics that we use in the present paper.
 
In Sec.~\ref{sec:dynamics} we study the dynamics of the constrained
system.  By fitting the value of the free energy for the disordered
state as a function of temperature and comparing it to the one of the
ordered state we first obtain an accurate estimate for the transition
temperature, which is in good agreement with the result from the CMFM.
We then show that there is no sign of the above mentioned
thermodynamic transition to the FMFS. The system instead becomes
supercooled and undergoes a lower-temperature non-equilibrium
transition from the supercooled liquid phase to a frozen
``polycrystalline'' phase.  The transition shows features that are
characteristic of first order phase transitions, such as a
hysteretic behavior as a function of temperature.  The underlying
physics behind this phenomenon is understood by studying the spin-spin
autocorrelation function as well as the evolution of the internal
energy and other physical quantities when we cool the system at
different cooling rates or after a quench from infinite temperature.


%
%
\section{
\label{sec:model}
The model and its possible experimental realizations}
In this section we review Baxter's 3-color model, and present two of
its possible experimental realizations in lattices of superconducting
devices in some detail. We show that the 3-color model and these two
realizations can be described as an Ising model on a hexagonal
lattice, with a plaquette constraint of $\pm 6, 0$ for the sum of the
spins around each hexagon. It is important to notice that while the
3-color model is only $\mathbb{Z}_2$ symmetric in the Ising spin
representation, the superconducting realizations have a larger
$\mathbb{Z}_2 \times U(1)$ symmetry due to the superconducting
phase. This difference is particularly relevant for the possible
defects that can originate in an allowed configuration and for their
dynamic behavior.

The one extra ingredient that we add to Baxter's 3-color model is a
local interaction. In the Ising spin representation, this interaction
takes the form of a nearest neighbor spin-spin interaction. It has the
effect, in the 3-color model, of aligning or not bonds of the same
color on neighboring sites. The extra interaction is responsible for
all the interesting thermodynamical and dynamical effects that are
studied in this paper. Moreover, in the lattices of superconducting
devices these interactions are always present.

\subsection{The three-color model}
\begin{figure}[!ht]
\centering
\includegraphics[width=0.9\columnwidth]{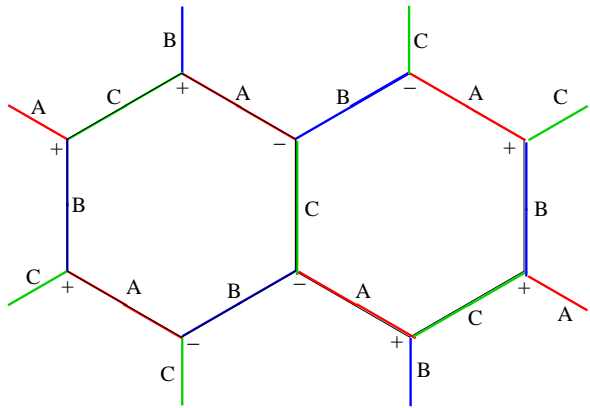}
\caption{
\label{Gluing}
The gluing of the ABC vertices gives Baxter's three coloring model on 
the hexagonal lattice. To every vertex we can associate a chirality spin 
depending on the order in which the three colors appear counterclockwise 
around the vertex: $+/-$ for even/odd permutations of the sequence ABC. 
}
\end{figure}

The three-color model consists of vertices having three bonds of different 
colors: A, B and C. These different colors can be thought of as three 
different phases differing pairwise by $\pm 2 \pi /3$, which is how we 
will later connect the model to arrays of superconducting devices. One can 
naturally associate to each vertex a chirality spin $\pm 1$ depending 
on the counterclockwise or clockwise ordering of the phases, as shown in 
Fig.~\ref{Gluing}. A hexagonal lattice is constructed with these 
vertices by connecting the bonds, where the connected bonds must share 
the same color. As we show below, the chirality spins cannot adopt an 
arbitrary configuration. Indeed, the spins must satisfy the constraint that 
their sum around any hexagon of the lattice is $\pm 6, 0$. 
On the other hand, given an allowed configuration of the spins, there 
are clearly three different corresponding color configurations, since 
any global even permutation of the colors in the lattice gives rise to 
the same spin configuration. 
In the absence of any kind of interaction this model corresponds to the 
Baxter's three coloring model on the hexagonal lattice. 
The partition function $Z$ has a 
purely entropic origin and its value is given by the number of ways of 
coloring the bonds of the hexagonal lattice. This number is known to 
grow exponentially with the system size. Indeed, Baxter solved 
exactly this model and showed that 
$Z = W^N$ for large values of the number of sites $N$, where 
$W = 1.2087...$ is the entropy per site~\cite{Baxter}. 

It is worth discussing in detail how the system can rearrange from one 
allowed configuration to another. 
No single-bond flip or double-bond exchange is allowed 
without violating the constraint in the neighboring vertices. 
However, we can notice that  by choosing one vertex and two colors, 
say A and B, we can uniquely define a loop by taking the sequence of 
ABAB... bonds starting from the chosen vertex. 
The loop must be non self-intersecting and closed, the last 
property holding only if the system has periodic boundary conditions. 
Clearly, if we pick one such loop and we flip the color sequence, say 
ABAB... to BABA..., the color constraint is preserved. These loop flips 
(or updates) provide a mechanism for the system to move around the phase 
space of allowed configurations. 
In Sec.~\ref{sec:defects} we will show how the loop updates originate 
from local constraint-violating defects. 
 
Notice that, given any allowed configuration, every vertex belongs to one 
and only one of such loops. Thus, by simply removing all the 
bonds of one of the three colors (say C), we realize one of the three 
possible simultaneous mappings of the system to a fully packed loop 
configuration on the hexagonal lattice 
which, at large scales, can be described by an $su(3)$ 
level 1 Wess-Zumino-Novikov-Witten (WZNW) model~\cite{Kondev}. 

The 3-color model becomes even richer when we introduce a nearest
neighbor spin-spin interaction in the Ising representation, which we
do in subsection \ref{sub:int}, after we discuss the experimental
realizations right below.

\subsection{The Josephson junction array of superconductors}
A possible experimental realization of the model is given by a
Josephson junction array of triangles of a superconductor with broken
time reversal symmetry. For example, there is experimental evidence of
a $p_x \pm i p_y$ order parameter in the compound
Sr$_2$RuO$_4$~\cite{Mackenzie,c-axis}; here the two possible states
$p_x \pm i p_y$ correspond to the chirality spin $\pm 1$ defined
above. The same geometry we propose here with $p \pm i p$ states has
also been studied by Moore and Lee, who in addition to the p-wave
states have also looked at $d\pm i d$ superconductors~\cite{Joel},
believed to be realized by the recently discovered hydrated cobalt
oxide compounds. In their work, they have also discussed other type of
arrays in triangular and square lattices.

\begin{figure}[!ht]
\centering
\includegraphics[width=0.9\columnwidth]{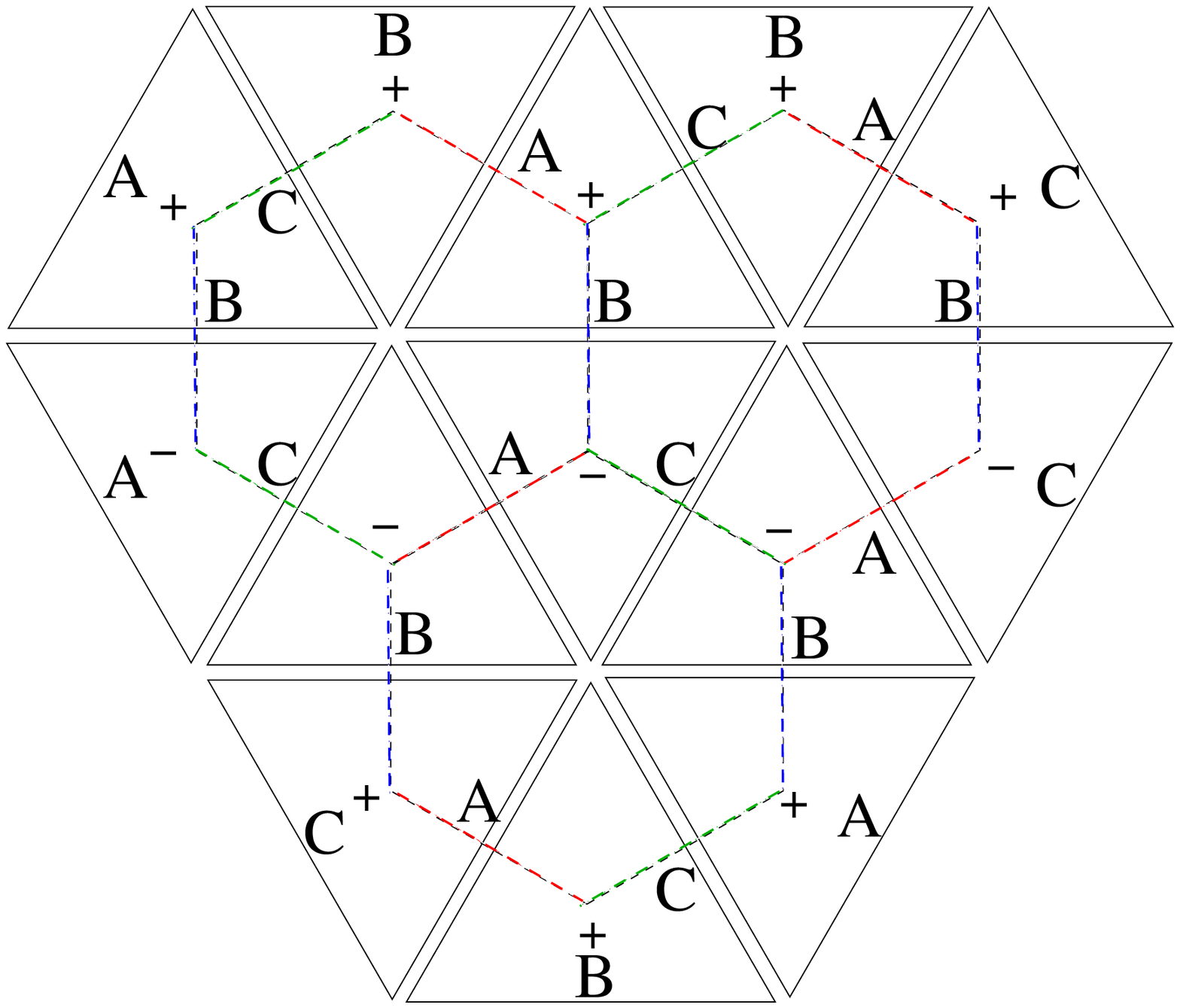}
\caption{
\label{Hextria}
An example of the correspondence between the Josephson junction array
and the three-color model, provided we identify the three colors with
the values of the phases of the order parameter in the middle of each
triangle edge.  Notice that ferromagnetic order among nearest
neighboring spins corresponds to aligning the bonds with the same
color along the same direction.  }
\end{figure}
\begin{figure}[!ht]
\centering
\includegraphics[width=0.75\columnwidth]{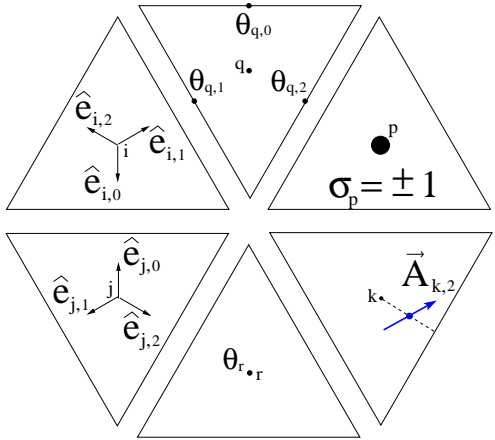}
\caption{
\label{triangle}
Labeling of the edges of the up and down triangles, with the relative 
unit vectors $\hat e_{i,a}$. 
While the chirality spins $\sigma_p$ sit at the centers of the triangles, 
the ``gauge'' fields (an example of which is shown in one of the triangles) 
sit at the midpoints of the segments joining the centers of the triangles 
to the corresponding edges. 
Examples of the $U(1)$ phase $\theta_r$ and of the edge phases 
$\theta_{q,a}$, $a=0,1,2$ are also shown. 
}
\end{figure}

In the $p_x \pm i p_y$ Josephson junction arrays, the three colors
correspond to the three relative phases of the order parameter in the
middle of each of the edges of the triangles, which differ by $\pm 2
\pi /3$ (see Fig.~\ref{Hextria}). (To be precise, the phase of 
the order parameters are defined in momentum space; but, as it can be
deduced from the analysis carried out in Appendix
\ref{AppB}, one can think in real space by considering the
phases for the momenta that point along the directions perpendicular
to the three faces of each triangle.)
The superconducting order parameter of each triangle has also an overall 
$U(1)$ degree of freedom. Therefore, at the center of each of its three 
edges, one can define a phase $\theta_{i,a}=\theta_i\pm\frac{2\pi}{3}a$ 
for the triangle at site $i$, along its $a$-th edge ($a=0,1,2$), where the 
edges are labeled counterclockwise starting from the horizontal one (see 
Fig.~\ref{triangle}). 
The $\pm$ sign corresponds to the chirality $\sigma_i=\pm 1$ of the
$p_x \pm i p_y$ state at site $i$. The Josephson coupling $-U
\cos(\theta_{i,a} - \theta_{j,a})$ along an edge shared by two
neighboring triangles tends to align the phases $\theta_{i,a}$ and
$\theta_{j,a}$. In the $U\to \infty$ limit one recovers Baxter's
3-coloring model, modulo a global $U(1)$ phase. Notice that, in this
infinite $U$ coupling limit, the only difference between this system
and the 3-color model (in the spin representation) described in the
previous subsection is a $\mathbb{Z}_2 \times U(1)$ symmetry instead
of a simple $\mathbb{Z}_2$ symmetry. We will show in
Sec.~\ref{sec:defects} how this difference allows for a wider variety
of defects in the Josephson junction array rather than in the three
coloring model.
\subsection{The \textit{kagom\'e} network of superconducting wires}
Another (related) realization of the 3-color model is given by a
superconducting
\textit{kagom\'e} wire network in the presence of a magnetic 
field~\cite{Higgins,Xiao,Park} such that the magnetic flux per
triangular plaquette is one half of a flux quantum ($f = 1/2$). Using
a Ginzburg-Landau analysis, Park and Huse~\cite{Park} showed that the
possible superconducting phases must have a gauge-invariant phase
change around each elementary triangle equal to $\pm \pi$, and a gauge
invariant phase change along each wire segment equal to $\pm \pi/3$.
They also show that the allowed minimum free energy states of this model 
are equivalent to ground states of the XY \textit{kagom\'e} antiferromagnet, 
which are in one-to-one correspondence to the three-color model 
configurations, modulo a $U(1)$ phase analogous to the one in the Josephson 
junction array. 
The $\pm 1$ chirality spin can be immediately read from the value of the 
(counterclockwise) phase change around each triangle $\pm \pi$, i.e. from 
the value of the induced flux through each triangle: $0$ or $1$ flux quantum. 
Even though this realization seems quite similar to the previous one, 
there are differences that arise mainly from the fact that time-reversal 
is explicitly broken by the external field in the wire networks. 
For example, the $\pm \pi$ chiralities do not have the same energy in 
the case of wires of finite width. We refer the reader to the thorough 
discussion of the energetics by Park and Huse~\cite{Park}. 
\subsection{Mapping to a constrained Ising model}
The hard constraint of the 3-color model imposes a hard constraint in the 
allowed configurations of the chirality $\pm 1$ Ising spins. Here we show 
that in the spin representation the hard constraint requires that any 
elementary hexagonal plaquette $P$ must have total magnetization: 
\beq
\sigma^{\mbox{\small\hexagon}}_P =
   \sum_{i \in P} \sigma_i =
   \pm 6,0.
\eeq
A similar result was obtained by Di Francesco and Guitter when
connecting the folding problem in the triangular lattice to the
3-coloring model~\cite{DiFrancesco1}. In our proof, we make use of
phases accumulated along paths on the hexagonal lattice, requiring
that these phases are single valued. This approach is more appropriate
to the discussion of superconducting systems and their defects
(integer and fractional vortices) that we present in this paper.

Indeed, as we show, one can obtain a simple interpretation of the hard 
constraint by identifying the accumulated phase around any loop lying on 
links of the hexagonal lattice with the circulation of a vector 
potential. For concreteness, we will use the example of the Josephson 
junction array in the discussion, but the argument is general. 

The phase $\theta_{i,a}$ on the edge $a$ of the superconducting triangle $i$ 
can be written as: 
\beq
\theta_{i,a}=\theta_i+\hat e_{i,a}\cdot\vec A_{i,a},
\eeq
where $\hat e_{i,a}$ is the unit vector that points from the center of 
triangle $i$ to its $a$-th edge, and the ``gauge'' potential $\vec A_{i,a}$ 
is defined at the center of such segment (see Fig.~\ref{triangle}). 

The phase difference across a face $a$ between triangles $i$ and $j$ is: 
%
\beq
\theta_{i,a}-\theta_{j,a}=\theta_i-\theta_j
+ [\hat e_{i,a}\cdot\vec A_{i,a} - \hat e_{j,a}\cdot\vec A_{j,a}]\ .
\eeq
%
The last term is simply the discrete sum equivalent of 
$\int d\vec\ell\cdot \vec A$ (notice that for neighboring sites $i,j$ 
the unit vectors are opposed, $\hat e_{i,a}=-\hat e_{j,a}$). 
 
Now recall that one can write 
$\theta_{i,a}=\theta_i+\frac{2\pi}{3}a\, \sigma_i$ and 
hence the vector potential is such that: 
\beq
\hat e_{i,a}\cdot\vec A_{i,a} =\frac{2\pi}{3} a\, \sigma_i
\eeq

\begin{figure}[!ht]
\centering
\includegraphics[width=0.4\columnwidth]{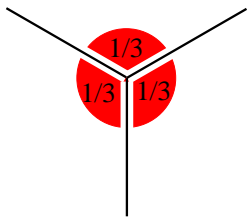}
\caption{
\label{frac-of-vortex} 
A vortex sitting at each vertex in the hexagonal lattice 
is shared by three hexagons. Hence, the contribution to an 
anti-clockwise accumulation of phase around a hexagon encloses one 
third of each of the six vortices sitting at the six vertices in 
the loop. 
}
\end{figure}

What is the corresponding magnetic field? This is simple to 
answer, by looking at the accumulated phase around a loop. 
Consider an elementary anti-clockwise hexagonal loop. The loop 
visits six triangles, and the portion of the loop within each 
triangle enters through face $a$ and exits through face $a-1$ (mod 
3), so that the accumulation of the vector potential along that 
portion of the loop is: 
\begin{eqnarray}
\hat e_{i,a-1}\cdot\vec A_{i,a-1}
-
\hat e_{i,a}\cdot\vec A_{i,a}
&&=\frac{2\pi}{3} (a-1)\, \sigma_i - \frac{2\pi}{3} a\, \sigma_i
\nonumber\\
&&=-\frac{2\pi}{3} \sigma_i \; . 
\label{contflux}
\end{eqnarray}
The above result, that each of the six sites visited by an 
elementary hexagon loop contributes $-\frac{2\pi}{3} \sigma_i$ to 
an anti-clockwise accumulation of phase around the loop, has a 
very simple interpretation. Each Ising spin $\sigma_i=\pm 1$ 
corresponds to a $\mp 2\pi$ vortex sitting at a vertex of the 
hexagonal lattice. Each vertex is shared by 3 hexagons; hence each 
hexagon can be thought to contain $1/3$ of that vortex, as 
depicted in Fig.~\ref{frac-of-vortex}. This is why the 
contribution from the hexagonal path going through vertex $i$ 
picks up the phase $-\frac{2\pi}{3} \sigma_i$ as shown above. 
Basically, the vortex is divided equally among the three 
neighboring hexagons sharing the common vertex. 

Using equation (\ref{contflux}) we can now compute the flux 
encircled by an elementary hexagon on plaquette $P$; it is given 
by: 
\beq
\Phi^{\mbox{\small\hexagon}}_P=
-2\pi/3\;   \sum_{i \in P} \sigma_i =
-2\pi/3\; \sigma^{\mbox{\small\hexagon}}_P
\eeq
Therefore the flux enclosed by an elementary hexagonal loop is just 
1/3 of the sum of the vorticities in the six sites. Now, matching the 
color scheme after going around any closed loop requires the phase 
around any hexagon to be uniquely defined (mod $2\pi$), which in turn
requires the flux to be a multiple of $2\pi$: $2\pi/3\;
\sigma^{\mbox{\small\hexagon}}_P = 0$ (mod $2\pi$), that is
$\sigma^{\mbox{\small\hexagon}}_P = \pm 6,0$ (notice that
$\sigma^{\mbox{\small\hexagon}}_P$ is even). Since the total flux
inside any loop is given by the sum of the fluxes through each
elementary hexagon, then the condition
$\sigma^{\mbox{\small\hexagon}}_P = \pm 6,0$ grants the
the phase to be uniquely defined (mod $2\pi$) around any loop.

Once the $\sigma^{\mbox{\small\hexagon}}_P=\pm 6,0$ constraint is
satisfied, there is a one-to-three mapping of any spin configuration
to a configuration of the color model, since there are three even
permutations of the colors that produce the same chirality spin
configuration.

In the case of the \textit{kagom\'e} wire networks at half-flux {\it per} 
triangle (or vertex of the hexagonal lattice), each triangle will 
accommodate either 0 or 1 vortex. So instead of $\sigma_i=\pm 1$ 
one has a variable $n_i=0,1$. Still, the vortices are split equally 
into three pieces, and the circulation around a hexagonal plaquette 
$P$ going through the centers of the \textit{kagom\'e} triangles is 
$\frac{2\pi}{3} N^{\mbox{\small\hexagon}}_P=\frac{2\pi}{3}\sum_{i \in P} n_i$. 
The circulation is a multiple of $2\pi$ if 
$N^{\mbox{\small\hexagon}}_P=6,3,0$. Indeed, the fact that the  
vortices in the elementary triangles are shared by three sites was 
used by Park and Huse~\cite{Park} in their argument for 
fractionalized vortices in the \textit{kagom\'e} superconducting 
wire networks. 

For finite $U$, there are defects that violate the 
$\sigma^{\mbox{\small\hexagon}}_P=\pm 6,0$ constraint; we shall discuss these 
defects in detail in section~\ref{sec:defects}, where we study integer and 
fractional vortices, as well as open segments of closed two-color loops. We 
analyze whether these different defects are confined or deconfined, and their 
importance in determining the ilk of the processes responsible for the 
dynamics. 
\subsection{Interactions}
\label{sub:int}
Each experimental realization of our model contains sub-dominant 
effects that may lead to a degeneracy lifting of the ground state. In 
this paper we concentrate on the effect produced by nearest-neighbor 
interactions between the chirality spins: 
\beq
H = -  \sum_{\langle i,j \rangle} J \sigma_i \sigma_j,
\label{Hamising}
\eeq
where the coupling $J$ depends on the microscopic details of the problem. 
Such a coupling can arise, for example, if one considers the 
higher order effects of having an extended Josephson junction barrier 
between two neighboring triangles in the array geometry. 
In the Appendix~\ref{AppB} we show how to derive the constants $U$ and $J$ 
from a microscopic Hamiltonian for the array of Josephson couplings and we 
discuss the conditions for having $U \gg J$. 
The sign of the $J$ coupling is positive in this case. 

This nearest-neighbor interaction leads, in the color language, to 
an aligning or anti-aligning interaction between the bonds, depending on 
the sign of the coupling constant $J$ as it can be easily seen with 
the help of Fig.~\ref{Hextria}. For $J$ 
positive, the spin interaction is ferromagnetic and the zero-temperature 
ground state (GS) of the system has all the bonds with the same 
color aligned in the same direction. We will refer to this 
translation invariant state as the FMFS state, or single crystal 
state. For $J$ negative, the spin interaction is antiferromagnetic and 
the zero-temperature GS of the system is a configuration where the six 
bonds in every hexagon form a sequence of only two alternating 
colors, which is simply the N\'eel order in the hexagonal lattice. 

In the following section, we discuss the thermodynamics of this 
system considering only the phase space of the configurations 
allowed by the ABC coloring constraint or, equivalently, by 
the $\sigma^{\mbox{\small\hexagon}}_P=\pm 6,0$ constraint. 

%
%
\section{\label{sec:thermodyn}Thermodynamics of the defect-free model} 
\subsection{Small $J$ and the CFT description}
Since the model without interactions can be described by a WZNW CFT, it is 
tempting to use this technique to analyze its behavior for small values of the 
spin-spin interaction. 

The first step is to represent the system by a height model (see Kondev 
\textit{et al.} for details~\cite{Kondev}). 
Flat configurations of this height model correspond to the different N\'eel 
states of the system. In terms of the colors there is a total of six of those 
configurations which are arranged to form an hexagonal lattice. 
The coarse grained version is described by two fields  $\vec{h} = (h_1,h_2)$ 
and a locking potential $V(\vec{h})$ that favors the fields to lie in one of 
the flat configurations; this potential has then the periodicity of the 
hexagonal lattice. The action reads: 
\beq
S = \int d^2x \left(\frac{\pi}{2} |\nabla \vec{h}|^2 + V(\vec{h}) \right) 
\eeq
In this language, the spin-spin interaction introduces a perturbation which 
is proportional to the "locking potential" since, depending on the sign of 
$J$, it favors or opposes the locking in one of the flat configurations. 
In the WZNW language, the locking potential can be written as a 
current-current perturbation of the underlying WZNW model~\cite{Kondev}. 

\begin{figure}[!ht]
\centering
\includegraphics[width=0.9\columnwidth,height=0.95\columnwidth]
                {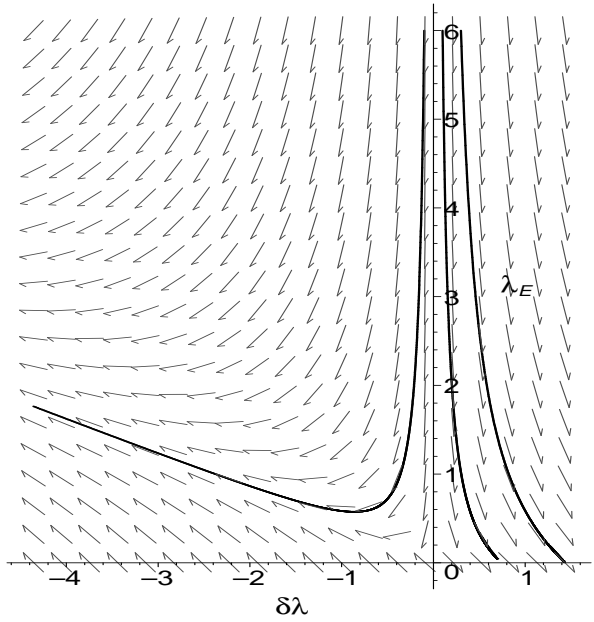}
\caption{
\label{RGflow}
Diagram of the RG flow for our model, where the horizontal axis corresponds 
to $\delta \lambda$ and the vertical to $\lambda_E$. The solid lines 
are numerical solutions of the system of equations (\ref{eqn::sysdiff}) 
for three different initial conditions, and are drawn for visualization 
purposes only. 
}
\end{figure}

When the spin-spin interaction is turned on, we can use this 
description to propose an action for the perturbed CFT. Since the A,B,C 
permutation symmetry is preserved, we can argue that the perturbing term to 
the pure CFT action should read: 
\bea
\int d^2x \left( \lambda_H (\sum_{i=1}^2 J^{H_i}_R J^{H_i}_L)+ \right. &&
\nonumber \\
&& \hspace{-0.25\columnwidth} 
\left. \lambda_E (\sum_{j=1}^3 J^{\alpha_j}_R J^{-\alpha_j}_L 
+ J^{-\alpha_j}_R J^{\alpha_j}_L )  \right)
\eea
where the $\alpha_j$'s are the generators of the root lattice of $su(3)$, 
and the Cartan generators $J^{H_i}$ are simply given by the derivatives of 
the height fields $\partial h_i$. 
The case $\lambda_E = \lambda_H$ corresponds to the $su(3)$ symmetric case. 
The one loop Renormalization Group (RG) equation in this case reads: 
\beq
\dot{\lambda} = - \frac{3}{2 \pi} \lambda^2 
\eeq
and for $\lambda>0$ the flow is toward the unperturbed level 1 $su(3)$ WZNW 
model, which can be identified with the $J=0$ case. 
In general, however, we just have the A,B,C permutation symmetry, and we 
cannot exclude the possibility of $\lambda_H \neq \lambda_E$. Defining 
$\delta \lambda = \lambda_H - \lambda_E $, the RG is now: 
\bea
\dot{\delta \lambda} &=& \frac{1}{\pi}~ \delta \lambda ~\lambda_E \nn  \\
\dot{\lambda}_E &=& - \frac{3}{2 \pi} \lambda_E^2  - \frac{1}{\pi} ~ \delta
\lambda ~\lambda_E,
\label{eqn::sysdiff}
\eea
where, at least for a small spin-spin interaction, we assume 
$| \delta \lambda | << \lambda_E$. The RG flow is as follows 
(see Fig.~\ref{RGflow}). For $\delta \lambda > 0$, the system flows to the 
line of fixed points $\lambda_E = 0$. While the $su(3)$ symmetry is broken, 
the system remains critical. We propose that this case corresponds to 
a ferromagnetic interaction, since it is equivalent to a decrease of 
the locking potential. 
This result is valid for small inter-spin couplings. As we show 
below, for large enough couplings a first order phase transition takes 
place. Since this is highly non-perturbative in the CFT language, this 
scenario is much better described by the Cluster Mean Field Method that we 
explain below. 
For an antiferromagnetic coupling, $\delta \lambda < 0$ and the flow
goes toward strong coupling, bringing the system off criticality and
forcing the system into anti-ferromagnetic ordering, as was argued by
Huse and Rutenberg~\cite{Huse} in their studies of the related
classical \textit{kagom\'e} XY model.
\subsection{The Cluster Mean Field Method: general approach} 
The CMFM is a technique that has proven to be very powerful in studying 
structural phase transitions in crystals and the thermodynamics 
of Vertex models \cite{Vaks}. 
When a system is constrained, fluctuations are considerably reduced and an 
appropriate mean field treatment can give very good results 
if the constraint is taken into account. The idea is to consider as the 
fundamental entity coupled to a ``molecular'' field, instead of a single 
spin, a cluster in which the allowed spin configurations are restricted 
by the constraint. The bigger the cluster, the more accurately fluctuations 
and constraints are taken into account. This method has given 
very precise results for the ice model \cite{Vaks} and is a good
candidate for giving an accurate picture of our constrained spin model 
in the hexagonal lattice. 

It is particularly simple to introduce the CMFM in the case of a corner 
sharing plaquette~\cite{Plaquette_note} lattice with Hamiltonian: 
\beq
H = \sum_{i,j}  J_{i,j} \sigma_i \sigma_j + h \sum_{i} \sigma_i,
\label{Genham}
\eeq
where the range of the $J_{i,j}$ interaction is shorter than the distance 
between the two farthest spins in a plaquette. This is the case for the 
present system. 
Let us assume that the lattice has $N$ spins and $2N/S$ plaquettes, where 
each plaquette has $S$ sites. The sums in the Hamiltonian can be 
rearranged as: 
\beq
H = \sum_{P} \left[ \sum_{i,j \in P} J_{i,j} \sigma_i \sigma_j + 
                    h \sum_{i \in P} \sigma_i \right] 
    - h \sum_{i} \sigma_i \, , 
\eeq
where the first sum is over all plaquettes $P$ and the last 
term compensates for the double-counting of the site energy term. 
The mean field approximation is obtained by considering each term as the 
sum over an elementary cluster (of $S$ and 1 spins respectively) coupled 
to an effective field representing the interaction with the rest of the 
lattice: 
\bea
H &\simeq& 
  \frac{2N}{S} \left[ \sum_{i,j \in P} J_{i,j} \sigma_i \sigma_j 
                    + (h + \phi_{ext}) \sum_{i \in P} \sigma_i \right] 
\nonumber \\
&& \hspace{4.5 cm} - N \left[ (h + \phi) \sigma_i \right]
\nonumber \\
&=& \frac{2N}{S} H_S - N H_1,
\eea
where $H_S$ and $H_1$ are the $S$- and $1$-spin cluster Hamiltonian 
respectively. Here, $\phi$ and $\phi_{ext}$ are proportional to the number 
of spins that are external to the cluster but connected to the internal spins. 
Since for the 1-spin clusters such number of external spins is twice the 
number for the $S$-spin clusters, we have $\phi = 2 \phi_{ext}$. 
Let us now define the effective internal energy per spin: 
\beq
\varepsilon = \frac{2}{S} \langle H_S \rangle_S - \langle H_1 \rangle_1,
\eeq
where $\langle \dots \rangle_S$ and $\langle \dots \rangle_1$ are the 
thermal averages computed with $H_S$ and $H_1$ respectively. Integrating 
then over the inverse temperature $\beta$ we get an effective free energy: 
\beq
\beta F = - \frac{2}{S} \ln{Z_S} + \ln{Z_1}, 
\label{EffF}
\eeq
where $Z_i = \textrm{Tr} \{ \exp(-\beta H_i) \}$, $i=S,1$, and the 
integration constant has been chosen such that in the case of 
unconstrained spins we get the trivial entropy $\ln(2)$ at infinite 
temperature. Minimizing the effective free energy with respect to $\phi$: 
\beq
\frac{\partial F}{\partial \phi} = 0
\label{minF}
\eeq
is equivalent to imposing the self-consistency equation for the magnetization: 
\beq
\langle \sigma \rangle_S = \langle \sigma \rangle_1
\eeq
and it gives us the optimal value for the field $\phi$, which determines 
the behavior of the system at a given temperature. 
An important benefit of this method is the fact that it can be 
extended to larger and larger clusters. This allows to improve 
systematically the accuracy of the results. 
\subsection{\label{CFMsec}Application of the CMFM to the defects-free model}
\begin{figure}[!ht]
\centering
\includegraphics[width=0.85\columnwidth]{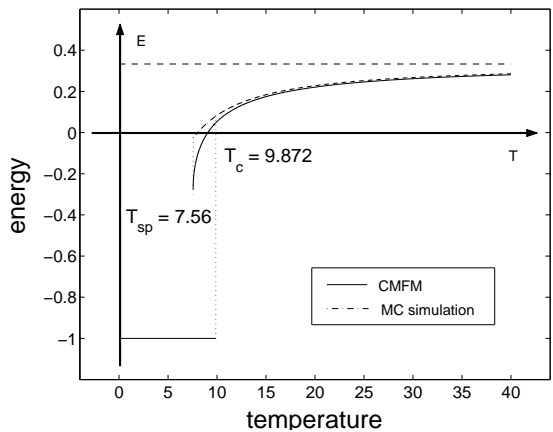}
\caption{
\label{energy} 
Plot of the internal energy per link as a function of the 
temperature. The solid line is the prediction from the CMFM and the
doted-dashed line is the result from the numerical simulation.  
}
\end{figure}

In order to be able to apply the CMFM to our problem in a straightforward 
way, it is convenient to switch to a bi-dual representation and describe 
our system in terms of spins $S_{ij} = \pm 1$ sitting on the links of the 
hexagonal lattice (see Figure~\ref{fig:frac-vortex}). 
These spins are given by the product of the original 
chirality spins $\sigma_i$ at the two vertices of each link: 
$S_{ij} = \sigma_i \sigma_j$. Obviously, the number of configurations of the 
$S$ spins is half the number of original $\sigma$ spin configurations, since 
we have quotiented by the global $\mathbb{Z}_2$ symmetry of the original 
model. The advantage of this mapping is that our lattice becomes now the 
(corner sharing hexagons) \textit{kagom\'e} net in which each spin $S_i$ is 
shared by two elementary plaquettes. 
In this description, the Hamiltonian (\ref{Hamising}) restricted to the 
nearest-neighbor interaction reads simply: 
\beq
H = -J \sum_{\alpha} S_{\alpha},
\eeq
where the index $\alpha$ refers to a link of the hexagonal lattice or 
a site of the bi-dual \textit{kagom\'e} lattice. 
The CMFM implementation is particularly easy since in this picture we just 
have an effective magnetic field $J$ in (\ref{Genham}). 
The clusters that we use are the single spin cluster and the elementary 
hexagon cluster (with 11 different configurations for the $S$ spins), and 
the corresponding partition functions are given by: 
\bea
Z_1 &=& ax^2 + 1/(ax^2) \nonumber \\
Z_6 &=& a^6 x^6 + a^{-6} x^{-6} + 3 (ax)^{2} +6/(ax)^2,
\label{Zdspins}
\eea
where $a=e^{\beta J}$ and $x=e^{\beta\phi /2}$. 
We can now obtain the values $\phi_{opt}$ corresponding to the minima of the 
effective free energy. Notice that $\phi_{opt}$ determines the equilibrium 
value of $\langle S \rangle$, i.e. of the internal energy per link of 
the original system. This method predicts the following scenario: 
for $T \rightarrow \infty$ we have $\langle S \rangle = 1/3$, which 
corresponds to an antiferromagnetic coupling in the system solely due to 
the constraint. 
This non-trivial value of the energy density is very close to the result 
obtained with the numerical method (see Sec.~\ref{sec:dynamics}). 
The cluster mean field method also gives a reasonable 
estimate for Baxter's entropy in the limit 
$T \rightarrow \infty$. Replacing (\ref{Zdspins}) in (\ref{EffF}) 
and taking the limit $T \rightarrow \infty$ we obtain the entropy per site 
$S = \ln(11/8)/2 \simeq 1.1726$, while the exact value is $1.2087...$. 
Since the analytical expressions for the forthcoming quantities are too 
cumbersome, we just mention here their numerical values. 
At $T \simeq 9.872 J $ the system undergoes a first order phase transition 
in which the energy density jumps from $\langle S \rangle \sim 0.05$ 
to a fully polarized state in which $\langle S \rangle$ is exactly $-1$ 
(see Figure~\ref{energy}). 
This transition has been first noticed via transfer matrix analysis 
by Di Francesco and Guitter~\cite{DiFrancesco2}  in the context of a 
folding transition. Our CMFM result is very close to their estimated 
critical temperature ($9.1 J$) and even closer to our Monte-Carlo 
estimate $9.6 J$ (see Sec.~\ref{sssec:tt_dyn}). 
In terms of the original spins, this 
behavior corresponds to the exotic scenario in which the magnetization 
jumps from $0$ to the fully saturated value $1$ at the critical point, as 
was argued by Di Francesco and Guitter~\cite{DiFrancesco2}. 
A similar kind of transition is also found in a frustrated spin model on the 
triangular lattice~\cite{Yin}, which turns out to be equivalent to a dimer 
model on the hexagonal lattice. Such kind of transition is accompanied  by 
slow dynamics and aging. As we will see below, slow dynamics is also a 
central issue in our case. 

Another temperature that we can compute via the CMFM is the spinodal 
temperature of the system. This is typical of first order phase transitions, 
where an appropriate fast cooling process can avoid crystallization and 
bring the system into a supercooled liquid phase. The spinodal temperature 
$T_{sp}$ is the temperature at which the supercooled liquid becomes unstable 
due to the crystal nucleation process. In this case, we can study the shape 
of the CMFM effective free energy as a function of $\phi$ for different 
temperatures. Starting from $T \sim \infty$ and lowering the temperature, 
the minimum corresponding to the liquid phase first becomes a local minimum 
(metastability) and eventually disappears. This metastability limit 
corresponds to the spinodal temperature $T_{sp} \simeq 7.56$ of the 
present model. 

The choice of the bi-dual spin representation to implement the CMFM is 
due to the fact that the system becomes a model for which the CMFM 
is particularly suitable. Indeed, in terms of the bi-dual spins, the system
becomes a \textit{kagom\'e} lattice seen as an array of corner sharing 
hexagons, in which now the new spins are sitting at the vertices. 
By associating to each of the 11 configurations for each hexagon its 
corresponding energy, the model can also be 
described as an 11 vertex model on the triangular lattice dual to the hexagonal.
This choice of variable usually limits the analysis since it does not allow 
to measure the magnetization of the system, which is the typical order 
parameter used to study phase transitions. 
In the present case however the energy density variable gives very good 
results in the characterization of the system since the transition is first 
order. For continuous phase transitions the situation is different. 
Even though the CMFM still gives a quite accurate result for the numerical 
value of the energy density (in contrast to the normal mean field method), 
it may fail in reproducing a subtle behavior such as an infinite slope point 
at $T_c$ in the energy vs temperature curve. 
In this case, measuring the magnetization of the system is a much more 
powerful tool to detect and study the second order phase transition. Thus, 
one needs to get back to the original spins instead of the bi-dual ones. 
Implementing the CMFM technique within the context of the real spins has 
two main disadvantages in our case. On one hand, the spins do not form corner 
sharing plaquettes, and relating the mean fields acting on the 1-spin cluster 
and on the 6-spin cluster becomes more difficult. On the other hand, since 
the coupling $J$ is a two spin nearest-neighbor interaction, a single 
variational mean field can not take simultaneously into account both the 
$J$ interaction (for which each spin interacts with its three neighbors) 
and the effective interaction due to the constraint (for which each spin 
interacts with all the 12 spins belonging to the three adjacent hexagons). 
\subsection{
\label{subsec:freen}
Free energy argument for a first-order phase transition}
The key point for understanding this particular phase transition is to 
understand the very peculiar nature of its FMFS ground state. 
As we already discussed before, in the FMFS state all the bonds of the 
same color are aligned in the same direction. As a result, any two-color 
loop is maximally straight and winds around the whole system. Thus, the 
smallest possible rearrangement of the FMFS configuration that produces 
another allowed configuration is the update of one of such loops. This 
is a striking feature of the ferromagnetic three-coloring model: the GS is 
separated from the first (1-loop) ``excited'' state by a system-spanning 
update which costs an energy: $E_{\textrm{1-loop}} - E_{\textrm{FMFS}} = 2JL$, 
where $E_{\textrm{FMFS}} = -3JL^2$ and $L$ is the system size ($2L^2$ sites, 
$3L^2$ bonds). 
Notice that if one prepares the system in the $T=0$ FMFS and starts to heat, 
the system is likely to remain in that state even for $T \rightarrow \infty$ 
for fast enough heating rates. 
Indeed, such an energy separation is likely to make the FMFS state 
metastable even for $T \rightarrow \infty$, in the thermodynamic limit. 
Since the FMFS state has zero entropy and the entropy of a 
straight winding loop is $\ln (3L)$, we can write the free energies of the 
two states: 
\bea
F_{\textrm{FMFS}} &=& -3JL^2 \nn \\
F_{\textrm{1-loop}} &=& -3JL^2 + 2JL - T \cdot \ln (3 L). 
\label{eqn_F_qsim0}
\eea
Clearly in the thermodynamic limit the energy cost $\Delta E \sim L$ 
overwhelms the entropic gain $\Delta S \sim \ln{L}$ and the excited state 
will never be favored over the FMFS state at any temperature. 
A similar argument applies to higher excited states, as long as their 
entropy is not exponential in the system size. 
The system is incapable (at equilibrium) to move out of its ground state in 
a ``smooth way''. In terms of configurations, it has to jump from a fully 
ordered state into a state with finite domain size. 
Since it is reasonable to assume that a finite-domain-size configuration 
has negligible magnetization, we can intuitively understand the origin of 
the complete first-order phase transition observed with the CMFM. 

The peculiarity of this transition and the relatively small variation of the 
internal energy in the disordered phase make it possible to obtain an 
estimate for the transition temperature by 
comparing the free energy of the FMFS configuration with the free energy 
of the disordered configuration. 
In order to compute the free energy of a disordered configuration, we use 
the average infinite-temperature internal energy of the system 
$E_{\infty} = JL^2$, an estimate derived via the CMFM in the previous 
section and confirmed by the numerical results (see 
Sec.~\ref{sec:dynamics}). Then, we can use Baxter's exact result for the 
residual entropy as an estimate of the entropy and obtain the free energy 
of a disordered state at all temperatures: 
\beq
F_{\textrm{disordered}} = JL^2 - T \cdot 2L^2 \ln{(1.2087)}.
\eeq
By comparing the free energy of the FMFS state $F_{\textrm{FMFS}} = -3JL^2$ 
with $F_{\textrm{disordered}}$ we obtain an estimate for the transition 
temperature $2J/(\ln{1.2087}) \simeq 10.55\,J$, which is reasonably close 
to the result from the CMFM $T_c \simeq 9.872J$. 

%
%
\section{\label{sec:defects}Defects and their role in the dynamics}
In this section we discuss the importance of defects in determining 
how the system can, dynamically, move from one of the allowed low 
energy configurations to another. 
For concreteness, let us start by discussing the Josephson junction 
arrays, i.e. the case of $\mathbb{Z}_2 \times U(1)$ symmetry. 
%
%
\subsection{Integer vortices}
For finite $U$, it is best to understand the system in terms 
of the chirality Ising spins, plus XY spin waves of the $U(1)$ sector. 
The lowest energy excitations over any configuration with Ising spins 
satisfying $\sigma^{\mbox{\small\hexagon}}_P=\pm 6,0$ are topologically 
trivial (no vortices) XY spin waves. 

When the $\sigma^{\mbox{\small\hexagon}}_P=\pm 6,0$ is preserved,
vortices of the $U(1)$ sector can only have vorticity that is an
integer multiple of $2\pi$. These vortices cost an energy of order 
of magnitude $U$, the vortex core energy. The $U(1)$ phase twist leads to 
the usual logarithmic interaction between a vortex/anti-vortex pair, 
\begin{equation}
\mathcal{E}_{1} \propto U\;{2\pi} \ln R \; .
\end{equation}
and these pairs are confined below a Kosterlitz-Thouless type
transition at a temperature scale $T_{KT}^{(1)}\propto U$. Since we are
interested in the regime of temperatures $T\ll U$ such that the
three-color constraint is enforced, these integer vortices will be
confined.

 
Now, what are the accessible excitations that break the 
$\sigma^{\mbox{\small\hexagon}}_P=\pm 6,0$ constraint? 
%
%
\subsection{Fractional vortices} 
\begin{figure}[!ht]
\centering
\includegraphics[width=0.9\columnwidth]{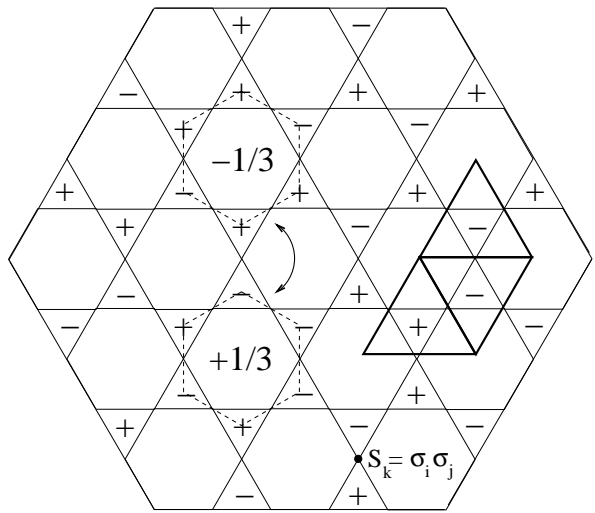}
\caption{
\label{fig:frac-vortex}
A pair of $\pm 1/3$-vortices created by a nearest-neighbor spin
exchange.  The solid-line lattice represents the \textit{kagom\'e}
network considered by Park and Huse~\cite{Park}. The Josephson
junction triangular array is represented instead by the bold
triangles.  The corresponding hexagonal lattice in our model is shown
only around the two defects (dashed line).  In the bottom right part
of the picture we show the mapping to the bi-dual representation used
in the CMFM.  }
\end{figure}

A fractional vortex excitation is illustrated in
Fig.~\ref{fig:frac-vortex}.  Such fractional vortices are always
created in pairs via a nearest-neighbor exchange of opposite pointing
spins and they have been discussed by Park and Huse~\cite{Park} in the
case of the superconducting \textit{kagom\'e} network.  A fractional
vortex excitation corresponds to a single hexagon that violates the
$\sigma^{\mbox{\small\hexagon}}_P=\pm 6,0$ constraint. We define its
fractional vorticity as
$\Gamma=2\pi\;\nu=\frac{2\pi}{3}\;\sigma^{\mbox{\small\hexagon}}_P$
(mod $2\pi$). Thus, we have $\nu=\pm1/3$ for
$\sigma^{\mbox{\small\hexagon}}_P=\mp2$ or
$\sigma^{\mbox{\small\hexagon}}_P=\pm 4$.

The presence of defects causes a fractional accumulation of the link sum of
the vector potential $\vec A_{i,a}$, the equivalent of $\oint d\vec\ell \cdot
\vec A$ in the continuum limit, that equals $\pm 2\pi/3$ (mod $2\pi$). Once
again it is useful to resort to the picture in Fig.~\ref{frac-of-vortex} to
understand that only one third of the vorticity associated to an Ising spin
at a vertex is included in the circulation around an elementary hexagon, and
hence the flux is $\frac{1}{3}\cdot 2\pi \sigma^{\mbox{\small\hexagon}}_P$.

To minimize the energy cost across the Josephson junctions, the
superconducting phases $\theta_i$ in the triangles must adjust accordingly to
pick this extra phase difference $\pm 2\pi/3$.  Hence, an excited state that
breaks the $\sigma^{\mbox{\small\hexagon}}_P=\pm 6,0$ constraint in the Ising
sector must be accompanied by a $U(1)$ phase twist that scales with the
distance $r$ from the defect as $1/(3r)$ (in units of the lattice spacing).

The $U(1)$ phase twist leads to a logarithmic interaction between a 
fractional vortex/anti-vortex pair a distance $R$ apart: 
\begin{equation}
\mathcal{E}_{1/3} \propto U\;\frac{2\pi}{3^2} \ln R \; .
\end{equation}
Thermodynamically, there is an entropic contribution to the free energy, 
which was calculated by Moore and Lee~\cite{Joel}, and shown to also be 
logarithmic. Therefore, there is a confining transition of the 
Kosterlitz-Thouless type at a temperature $T_{KT}^{(1/3)}\propto U/9$. If the 
Josephson coupling $U$ is large compared to the temperature $T$, which is the 
regime we are interested in, then one is deep in the confined phase, and 
fractional vortices are rather ineffective as a source of phase space 
reconfigurations. 

\subsection{Open segments of closed two-color loops}
There is a special way to flip Ising spins along certain strings lying 
on the hexagonal lattice that, while violating the 
$\sigma^{\mbox{\small\hexagon}}_P=\pm 6,0$ constraint, only costs 
energy at the extremities of the string, irrespective of its length. 

To understand these excitations, let us start by looking at the simple 
case of a single spin flip that violates the constraint on three 
neighboring hexagons. In terms of the color model, all colors remain 
perfectly well defined, with the exception of the one vertex where the 
spin flip occurred. The energy cost of this defect is of order $U$. It 
is possible that locally adjusting the $U(1)$ phase near the defect 
might slightly relieve this cost, but we have not investigated this 
issue. A single spin flip could split into a $+1/3$ and $+2/3$ (or 
equivalently, a $-1/3$) fractional vortex pair. These, however, are 
confined together at low temperatures compared to $U$, as we argued 
above. 

\begin{figure}[!ht]
\centering
\includegraphics[width=0.85\columnwidth]{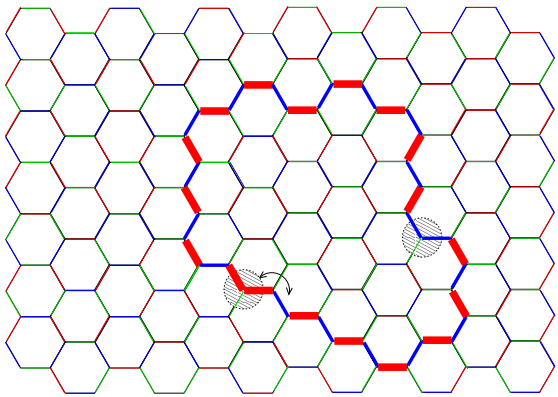}
\caption{
\label{fig:open_loop}
Defect pair at the end-points of an open string, with end-points
highlighted (shaded circles) and the relative two-color path (bold
links) shown in a configuration of the 3-color representation of the
model. The end-points can travel freely along the path via
nearest-neighbor color exchanges, such as the one outlined by the
double arrow. Eventually, the two end-points recombine by either
exchanging all the bonds along the path, or by leaving them all
unchanged.  }
\end{figure}

In the three-coloring model, this spin flip defect corresponds to the 
initial step of creating an open segment defect described 
hereafter. Out of the three bonds departing from the spin flipped 
site, two must have exchanged color (in order to change the chirality of 
the vertex), thus violating the color matching with the corresponding 
two neighboring sites. If we now move these two color 
defects starting from the two neighboring sites and performing the 
same original color exchange, we can propagate the defects at zero energy 
cost along a predefined path. Indeed, every color exchange will fix the 
previous color mismatch and create a new one, one lattice spacing apart. 
Notice that this process will flip all the spins between the two end points 
along the path. 
It is useful to recall the color description of the allowed 
low energy states. Imagine one follows an ABAB... sequence, that 
always forms a closed loop in an allowed configuration. 
We have already seen that flipping the whole loop to BABA...  maintains 
the system in an allowed configuration. It is also trivial to show that 
this update flips all Ising spins visited by the loop. 
While this is a rather non-local move, starting from a single spin flip 
(color exchange) and propagating the color defects as above, we can realize 
this move through a sequence of local updates. 
Instead of flipping the whole loop at once, one can do it in steps, 
flipping the spins along a piece of the loop sequentially. Notice that 
the energy cost of this string is paid only at the end-points and is 
of order $U$, as long as the sequence of spin flips moves on its 
two-color track. The end-points can be thought of as a defect pair 
connected by a string. This special path is hidden in the 
constrained Ising representation, but is clear in the 3-color one (see 
Fig.~\ref{fig:open_loop}). 
The defect pair, once formed, can diffuse around the one-dimensional 
loop, and it has two channels to decay back into an allowed state: 
either the defects recombine by going around the whole loop, leading 
to the BABA...  configuration, or they recombine without winding 
around the loop back to the original ABAB... configuration. These are 
the defects considered by Kondev \textit{et al.}~\cite{Kondev}. In 
the CFT description, they correspond to vertex operators with 
conformal dimension $1/2$. While, as we mentioned, for a fixed 
configuration of colors there is no confining force between pairs, an 
effective interaction appears because of entropic reasons, producing an 
algebraic decay with the separation distance for the partition 
function in the presence of such defects. However, for the dynamics 
one is really interested in the cost for a given 
configuration. Therefore, the formation and recombination of these 
defect pairs constitute the main mechanism responsible for the 
dynamical evolution of the system. 
 
The defect formation time just enters as an overall rescaling of the time 
steps for loop updates. Also, since the time it takes for the defects to move 
diffusively around the 1-D loop is algebraic in the loop length (and not 
exponential) we can neglect this correction and simply treat the whole loop 
update as a non-local elementary move, now with a justified local origin. 

%
%
\section{\label{sec:dynamics}Dynamics}
In order to study the dynamic properties of the system, we use Monte 
Carlo (MC) simulation techniques of an $N = 2L^2$-site hexagonal 
lattice ($3L^2$ bonds) with periodic boundary conditions. 
As we discussed in Sec.~\ref{sec:model}, the choice of the single-step 
update is non-trivial due to the color constraint. 
In Sec.~\ref{sec:defects} we argued that the 
open segments of closed two-color loops are the main actors in the 
dynamical evolution of the system, based on energy and confinement 
considerations. Thus, without loss of generality, we consider only loop 
updates as single-step updates of our MC technique. We also assume that 
the rate of formation of the open segment defects is low enough not to 
allow for defect proliferation (i.e. for the intersection of two 
different open segments before they recombine). 

To implement a loop update we proceed as follows: we first choose one 
site and two colors at random; then we compute the energy difference in 
the system for the update of the corresponding loop; eventually we accept 
or reject the update based on the usual Boltzmann probability. 
Notice that, with this choice of the single MC step, the update of a loop 
takes one unit of time, independent of its length. In a possible 
experimental realization we expect the two ends of an open segment defect 
to walk randomly along the corresponding closed path, until they 
recombine. Thus, our MC dynamics is accelerated and the rescaling of our 
MC time with respect to a possible ``real'' time is highly non-trivial. 
Since we are interested in studying the slowing down and freezing of the 
dynamics in the three coloring model, we choose to use the accelerated loop 
dynamics in order to be able to sample much longer time scales, otherwise 
inaccessible with a realistic update mechanism based on defect formation 
and recombination. 

In terms of the loops, one can notice that the two ordered 
configurations FMFS and N\'eel (ferromagnetic and 
antiferromagnetic respectively) correspond to the two extrema in loop 
curvature. In the FMFS configuration, the loops are completely straight 
loops, winding around the whole system. In the N\'eel 
configuration, the loops are maximally curved into single hexagon loops. 
For these reasons, we expect an entropic jamming in the approach to the 
FMFS state, for a ferromagnetic choice ($J>0$) of the interaction, as 
discussed in the case of infinite-range interactions by 
Chakraborty \textit{et al.}~\cite{Chakraborty}. 
Indeed, entropy favors rough and entangled loops, which in the 
infinite temperature limit have a fractal dimension equal to 
$1.5$~\cite{Kondev,Kondev2}. This creates a phase-space bottleneck 
due to the small number of configurations that allow the system to reach 
the FMFS state with straight, packed loops. 
On the other hand, the approach to the N\'eel state 
in the antiferromagnetic interaction case ($J<0$) is much smoother for the 
system. Even though this state has zero entropy by itself, 
single-hexagon flips allow the system to achieve a gain in entropy 
of the order of $\ln{L^2}$ with an energy cost of the order of $6J$. 
Indeed the N\'eel state corresponds to the 
\textit{ideal states} defined by Kondev and Henley~\cite{Kondev}, which 
have maximum entropy density in the sense that they allow for a maximum 
number of local rearrangements of the spins in accord with the constraint. 
Thus, we do not expect any jamming phenomena to play a role in this case. 

In this section we consider only the case of ferromagnetic interactions 
and we set $J=1$ as the unit of measure of energies and temperatures. 
In order to be able to access large simulation times, we choose the 
smallest system size for which our results do not show a significant 
dependence on system size ($L=18$). 
\subsection{Transition temperatures}
\subsubsection{
\label{sssec:tt_dyn}
Estimate of the thermodynamic transition temperature}
\begin{figure*}[!ht]
\begin{center}
\subfigure[\label{evolution(a)}
$m = 0.08$, $t = 1$ MC step]
{\includegraphics[width=0.74\columnwidth]{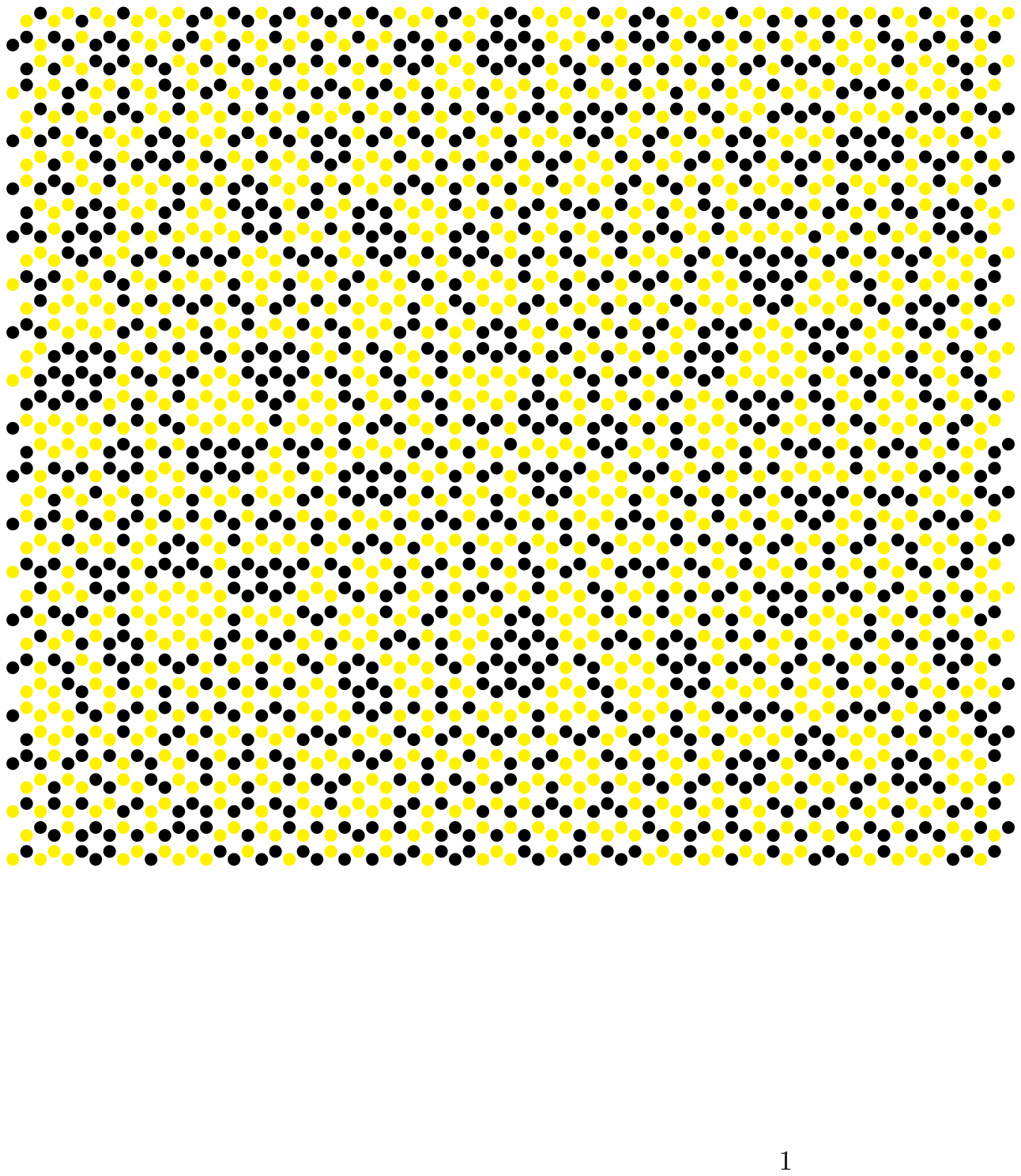}}
\hspace{1.5 cm}
\subfigure[\label{evolution(b)}
$m = 0.24$, $t = 28$ MC steps]
{\includegraphics[width=0.74\columnwidth]{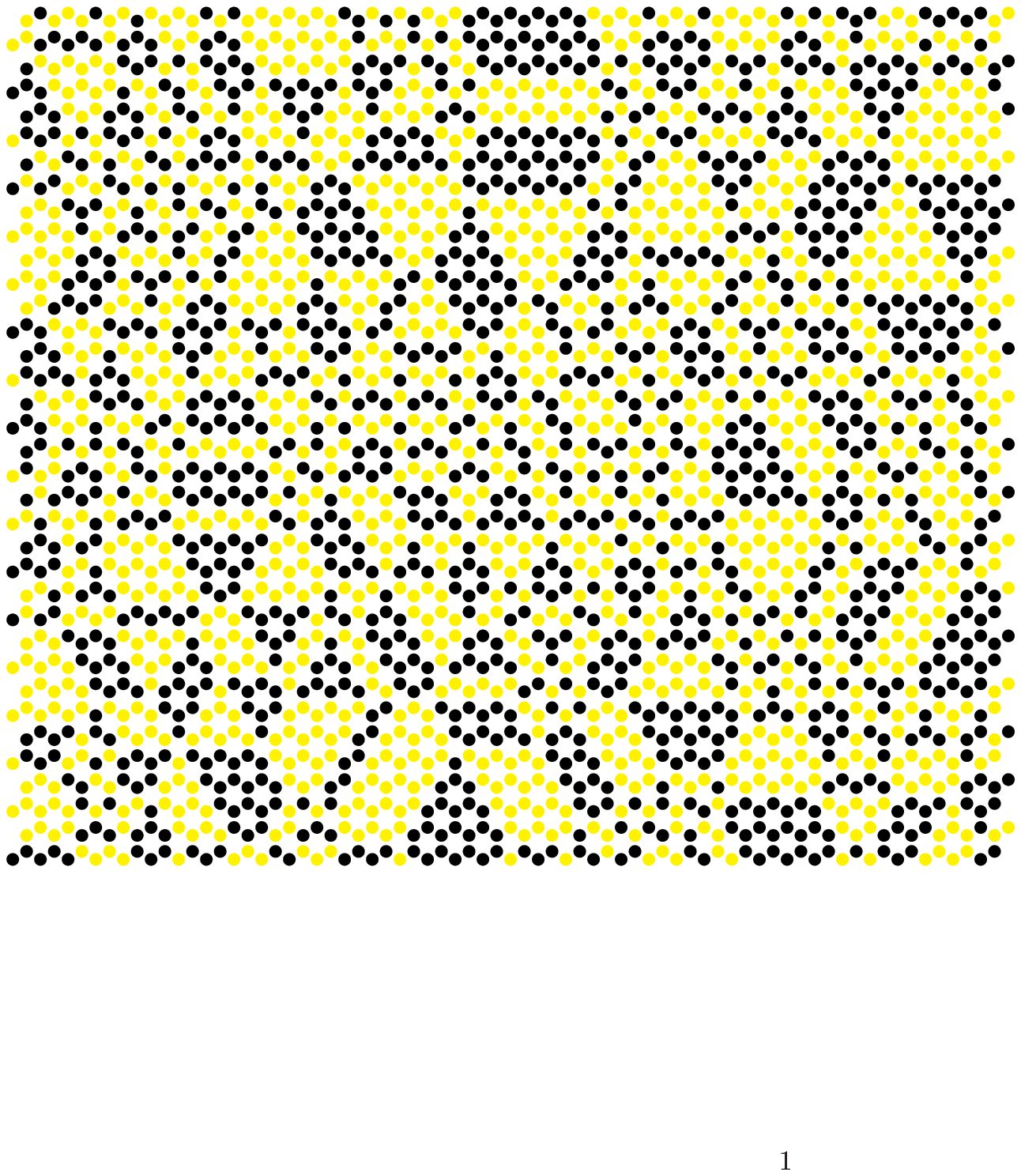}}
\\
\subfigure[\label{evolution(c)}
$m = 0.32$, $t = 49$ MC steps]
{\includegraphics[width=0.74\columnwidth]{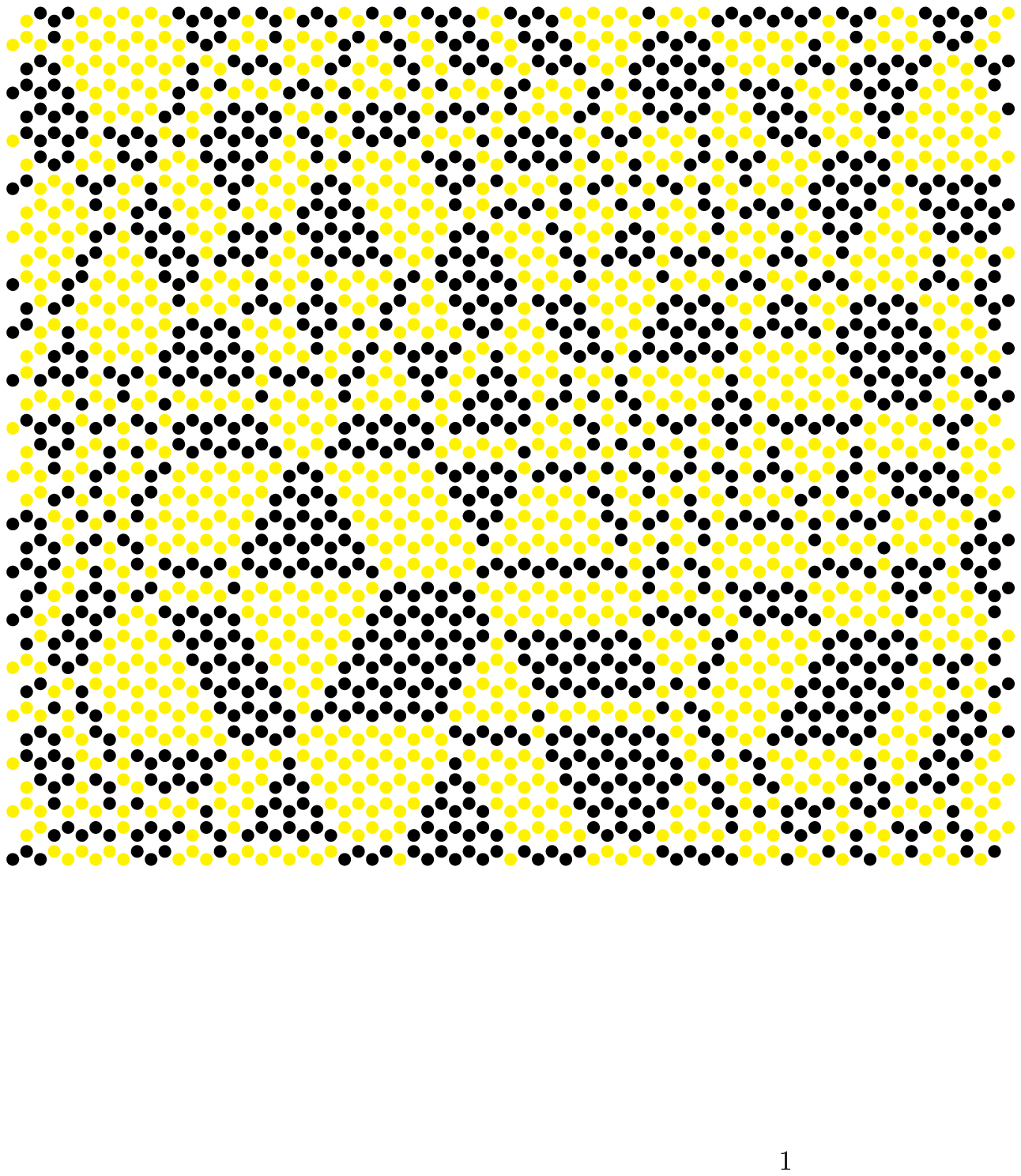}}
\hspace{1.5 cm}
\subfigure[\label{evolution(d)}
$m = 0.50$, $t = 192$ MC steps]
{\includegraphics[width=0.74\columnwidth]{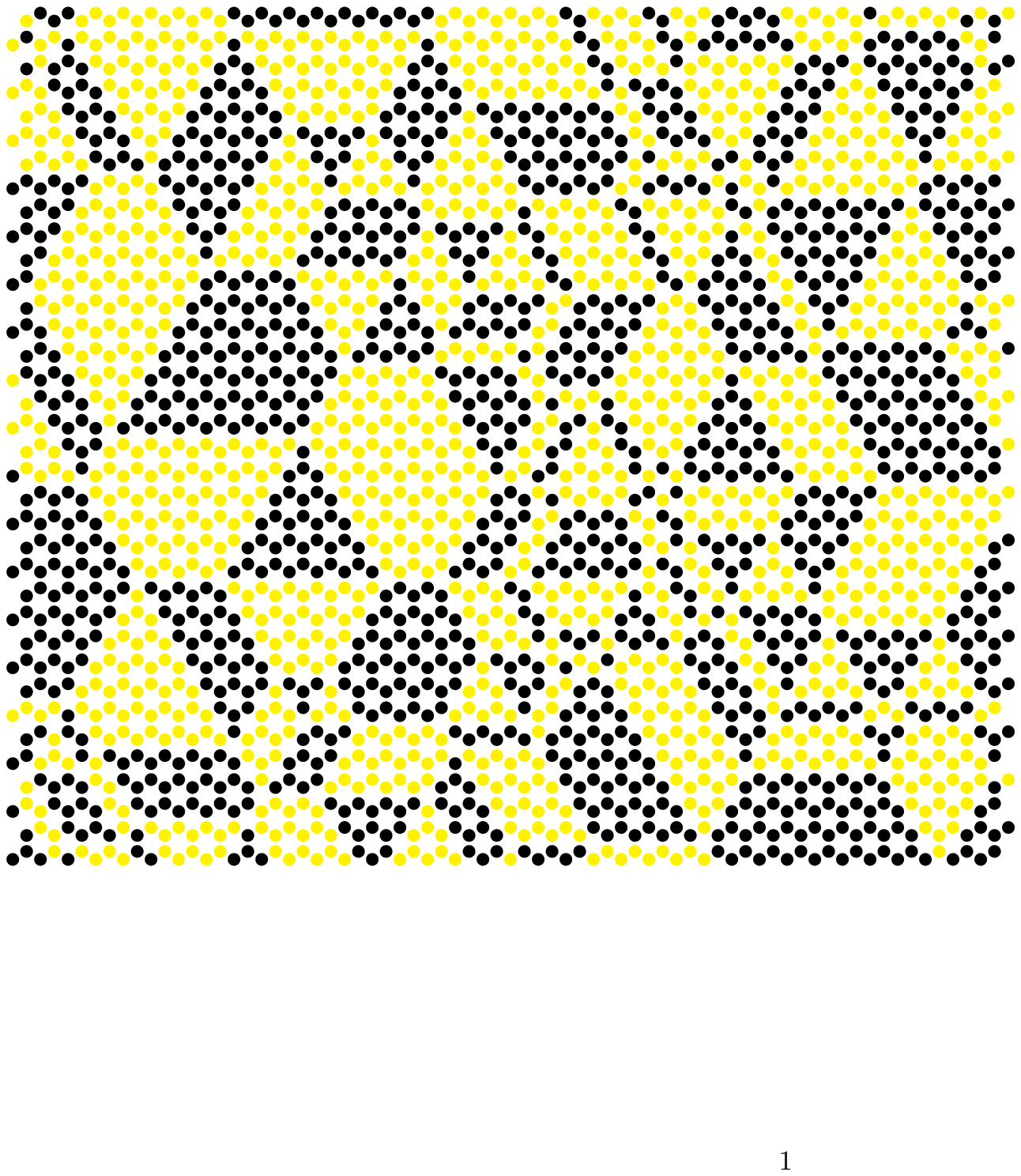}}
\\
\subfigure[\label{evolution(e)}
$m = 0.68$, $t = 5.7 \cdot 10^4$ MC steps]
{\includegraphics[width=0.74\columnwidth]{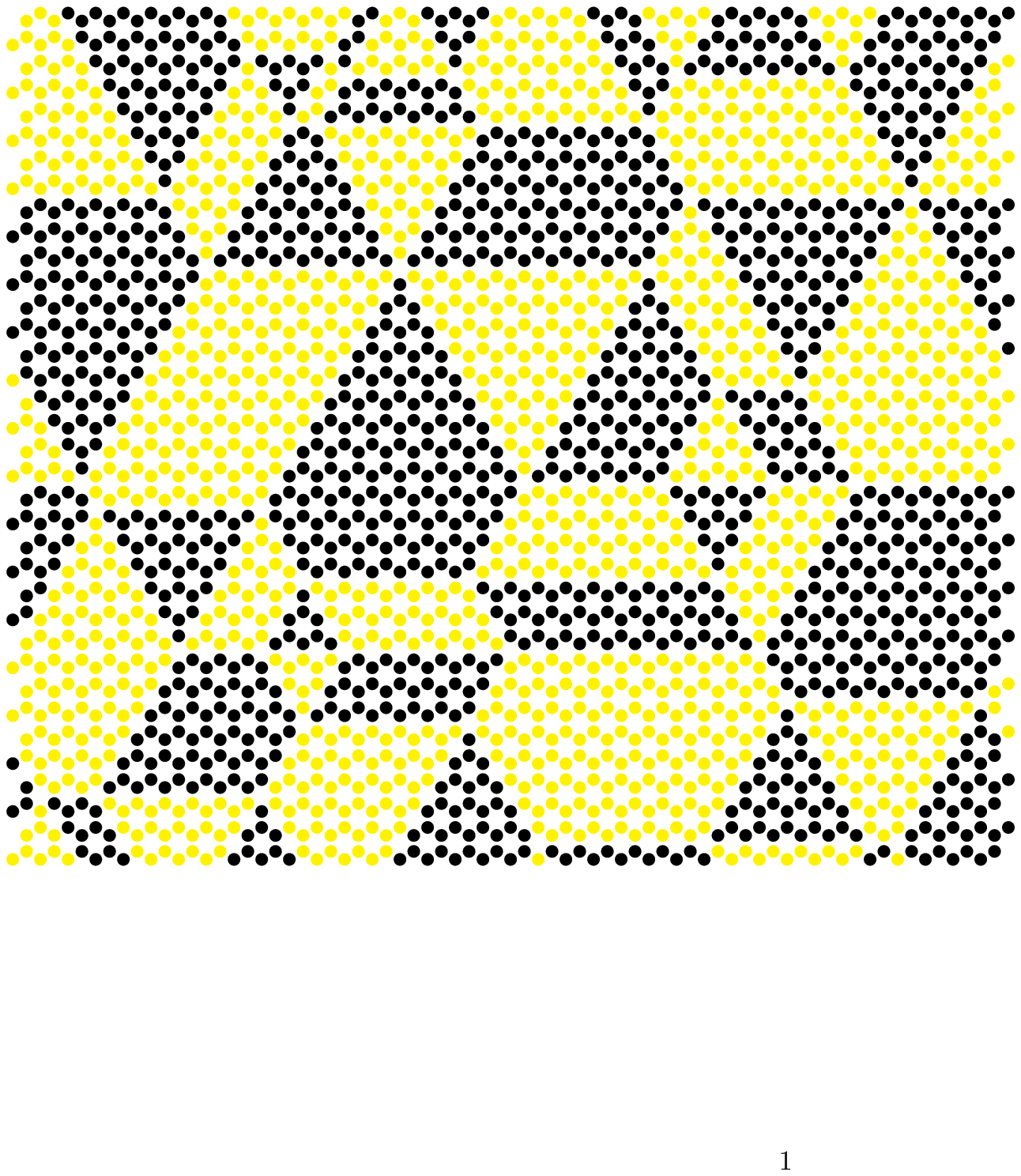}}
\hspace{1.5 cm}
\subfigure[\label{evolution(f)}
$m = 0.73$, $t = 5.4 \cdot 10^5$ MC steps]
{\includegraphics[width=0.74\columnwidth]{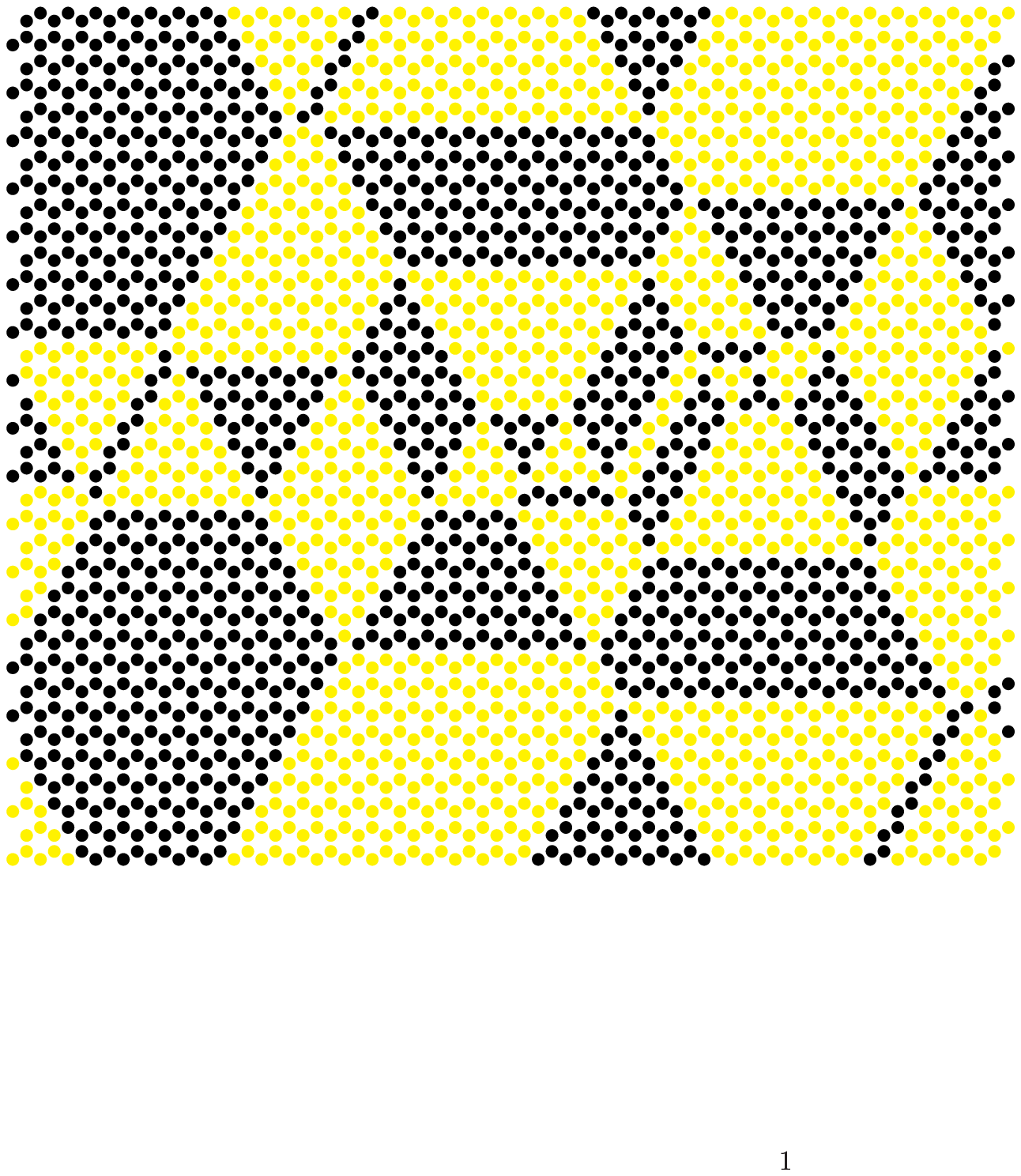}}
\end{center}
\caption{
\label{evolution}
Time evolution snapshots of the system after a quench from 
$T \sim \infty$ to $T = 6.0$ (at time $t=0$) below the transition 
temperature $T^* \simeq 8.1$. 
The dots represent the $2L^2$ vertices of the hexagonal lattice ($L = 36$) 
and the two colors correspond to the two values of the chirality spin. 
The lattice is wrapped along the horizontal axis and along the 
$60^\textrm{o}$ axis rotated counterclockwise above the horizontal. 
For each configuration, we report the measured crystalline mass $m$ and 
the time $t$ from the temperature quench. 
}
\end{figure*}

The first result that we observe both in cooling/heating 
simulations and in quenching simulations is the phase space 
``isolation'' of the single-crystal phase or FMFS. Even though at 
equilibrium the system must eventually favor the FMFS, we were 
unable to reach it within any simulation time, up 
to $10^7$ MC steps. The system prefers to settle into a frozen 
polycrystalline (P-xtal) phase with zero or close to zero average 
magnetization, and with very slow, event-dominated dynamics. 
In Fig.~\ref{evolution} we show the time evolution of the system after 
a quench in temperature from $T \sim \infty$ to $T = 6.0$. 
After a single MC iteration (Fig.~\ref{evolution(a)}), only a few small 
crystalline seeds are visible in a disordered liquid background. 
These seeds quickly develop into well-defined domains 
(Fig.~\ref{evolution(b)}), whose size grows with time until the system 
becomes frozen into the polycrystalline (P-xtal) phase 
(Fig.~\ref{evolution(d)}). Notice the domain boundaries 
following the ``crystalline planes'' of the hexagonal lattice in the 
polycrystal. 
The dependence of the crystalline mass $m$ on time $t$ reflects the 
remarkable slowing down in the dynamics once the system enters the 
polycrystalline phase. 

Even melting simulations starting from the FMFS phase and 
increasing the temperature are not useful to estimate the transition 
temperature. Indeed, they result in a large overestimate of $T_c$, since 
the melting time remains much larger than the simulation time well above 
$T_c$. 
 
\begin{figure*}[!ht]
\begin{center}
\subfigure[\label{Tinf_picture(a)}Original spins]
{\includegraphics[width=0.8\columnwidth]{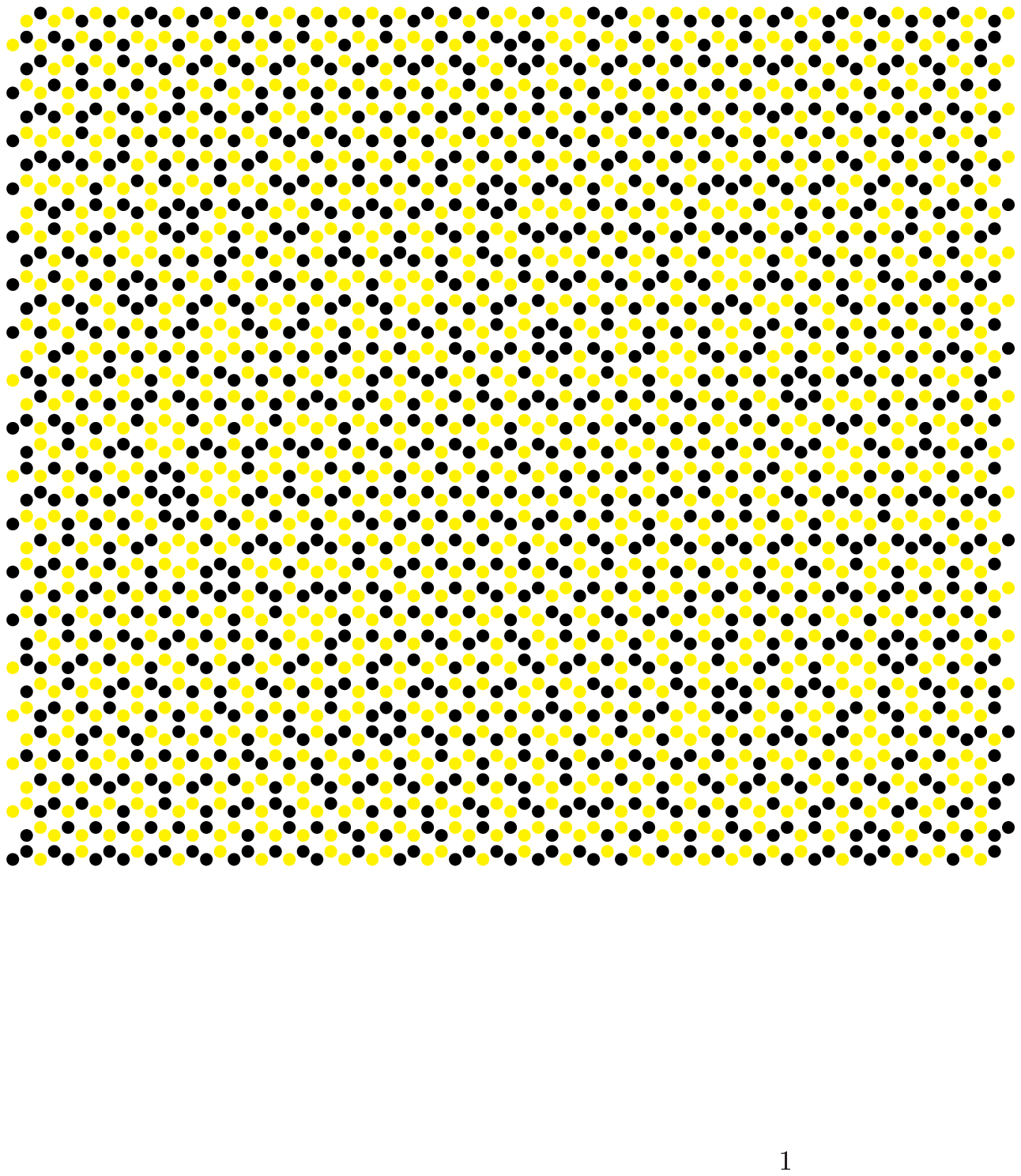}}
\hspace{1.5 cm}
\subfigure[\label{Tinf_picture(b)}
Same spins after a one-sublattice spin-flip operation]
{\includegraphics[width=0.8\columnwidth]{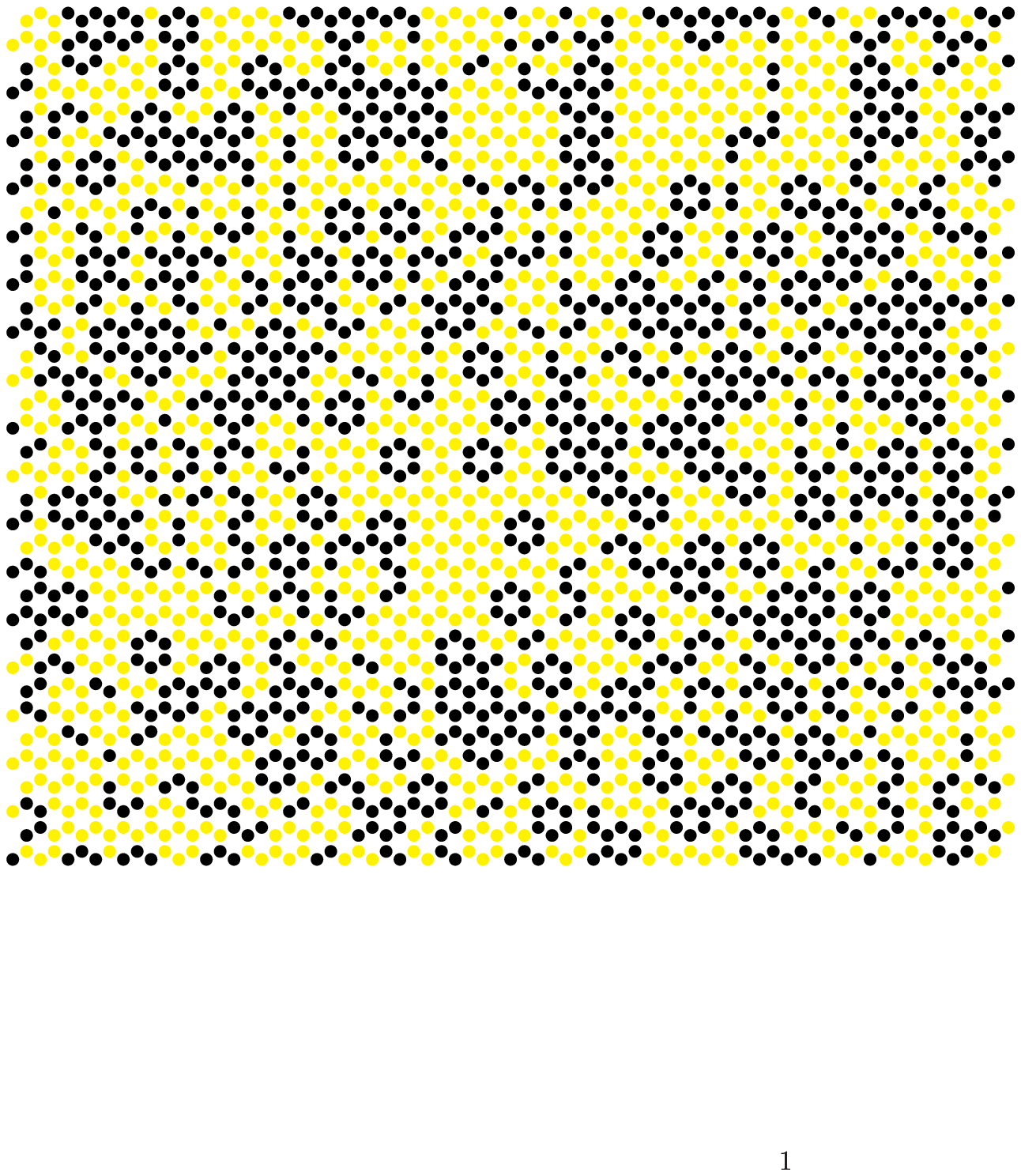}}
\end{center}
\caption{
\label{Tinf_picture}
A picture of a $L = 18$ system at infinite temperature: (a) the 
original chirality spins are shown; (b) we performed a spin-flip 
operation on one of the two sublattices in order to make visible 
the antiferromagnetic correlations due to the constraint. 
}
\end{figure*}

The only measure we can achieve of the thermodynamic transition 
temperature $T_c$ is by computing the free energy in the liquid 
and crystal phases by integration of the internal energy. For a 
single crystal we know that $f_{\textrm{FMFS}}=-1$ at all 
temperatures, where $f=F/(3L^2)$ is the free energy per bond. For 
the liquid phase, we use the curves in Fig.~\ref{EvsT_all} 
showing the dependence of the internal energy on the temperature. 
Notice that the asymptotic value of the internal energy at 
infinite temperature is different than zero. This is purely due to 
the constraint, which appears to be slightly antiferromagnetic in 
nature. A simple way to visualize this effect is to look at an 
infinite temperature configuration after performing a spin-flip 
operation on one of the two sublattices of the hexagonal lattice. 
The result is shown in Fig.~\ref{Tinf_picture}. 

An appropriate fit of the common high-temperature region of the internal 
energy (per bond) curves~\cite{fit_note}: 
\beq
\mathcal{E}_{\textrm{liquid}}(T)=c-a/T^b 
\label{eqn_Elq}
\eeq
gives $a \simeq 4.3$, $b \simeq 1.22$ and $c \simeq 0.336$. 
Notice that a naive high temperature expansion in powers of 
$\frac{1}{T}$ may be plagued by the criticality at hight temperatures. 
In this sense the nontrivial exponent $b$ may have an interpretation in 
terms of the CFT description at $ T \rightarrow \infty$. We can then 
integrate to obtain the free energy: 
\beq
\beta f(\beta) =
   \beta_0 f(\beta_0) +
   \int_{\beta_0}^{\beta}\textrm{d}\beta' \mathcal{E}(\beta');
\eeq
setting $\beta_0 = 0$ for the liquid phase and using the known residual 
entropy of the system, we obtain: 
\beq
f_{\textrm{liquid}}(T)=-\frac{2}{3}\ln{(1.2087)}T + c -
                        \frac{a}{(b+1)T^b},
\label{eqn_Flq}
\eeq
where the $2/3$ factor in front of the residual entropy comes from the fact 
that there are $3$ bonds every $2$ spins. 
Setting $f_{\textrm{liquid}}(T) = f_{\textrm{FMFS}} = -1$ gives the melting 
temperature $T_c = 9.6$, in good agreement with the results from the 
CMF method. 

Even though $T_c$ is the actual thermodynamic transition 
temperature, we are unable to observe this transition due to the 
incredibly large time scales involved in the approach to the FMFS 
state. As it appears from the results below, the system seems to 
be completely unable to sample the phase space region 
corresponding to the crystalline phase, at least on our simulation 
time scales, and it is confined to an ``effective phase space''. 
\subsubsection{The dynamic freezing transition}
Instead of going through the thermodynamic transition, the system 
remains in a supercooled liquid state below $T_c$, until it 
reaches a temperature $T^*$ where it evolves into a frozen 
polycrystalline state. 

Looking at Fig.~\ref{evolution(f)}, we can clearly see that the 
polycristallization is complete, in the sense that the domain 
boundaries are fully one-dimensional, with almost no interstitial 
liquid left. While the size of these domains increases with longer 
waiting times, the growth becomes extremely slow, basically 
stopped within our Monte Carlo time scales before reaching the 
single crystal configuration. This can be observed, for example, 
in the behavior of the zero-temperature saturation value of the 
energy in Fig.~\ref{EvsT_cool} and in Fig.~\ref{EvsT_plateau}. 
The energy is in fact a measure of the 
area-to-perimeter ratio in the polycrystalline phase, 
provided complete polycrystallization has been achieved. This is 
clearly the case in the $T \rightarrow 0$ plateaus in 
Fig.~\ref{EvsT_all}. Instead of approaching the value $-1$, 
characteristic of the FMFS state, these plateaux seem to approach 
a limiting value $\mathcal{E}^{\textrm{P-xtal}}(T=0) \sim -0.74$ 
for larger cooling times. 

The transition at $T^*$ can be seen as a dynamic phase transition and 
does not have a thermodynamic origin. However, we can reasonably 
establish a correspondence of this transition to a ``true'' 
thermodynamic phase transition in a related, more constrained 
system. As we show with the following analysis, the origin of the 
dynamic transition at $T^*$ resides in a free energy barrier that 
prevents the system from visiting a phase space region 
around the FMFS phase, at least within our simulation timescales. 
Since only winding loop updates can change the number of bonds per 
color per direction, it is possible to divide the phase space into 
topologically separated sectors by forbidding the update of 
winding loops. The FMFS configuration would then be in a 
topological sector by itself, and starting from an infinite 
temperature configuration with equal number of bonds per color per 
direction it would be impossible for the system to reach its 
natural ground state. With this constraint, the system is expected to 
show a phase transition into a state which is not the FMFS, with a 
behavior analogous to the one observed in the present model. 

This polycrystal transition is an intrinsic transition of the 
supercooled liquid phase, which would not exist in the infinite 
time limit. If we were able to wait infinite simulation times, we 
expect the dynamic transition at $T^*$ to disappear, replaced by 
the equilibrium transition at $T_c > T^*$. 

Since we cannot apply the same technique used above for $T_c$ to 
the polycrystalline state, we have to measure $T^*$ with a somehow 
more empirical method. We first prepare the system into an almost 
completely polycrystallized state by cooling it at very low rates. 
We then chose a particular value for the temperature $T$ and let 
it evolve in time. If it eventually reaches the liquid state, then 
we conclude that $T>T^*$; conversely if it completes the 
polycrystallization process. The choice of the initial state 
closer to the polycrystalline state rather than to the liquid one is 
merely due to the stronger metastability of the liquid phase, as 
it appears from the asymmetry in the hysteretic process with 
respect to $T^*$ (see Fig.~\ref{EvsT_hist}). 
\begin{figure}[!ht]
\centering
\includegraphics[width=0.85\columnwidth]{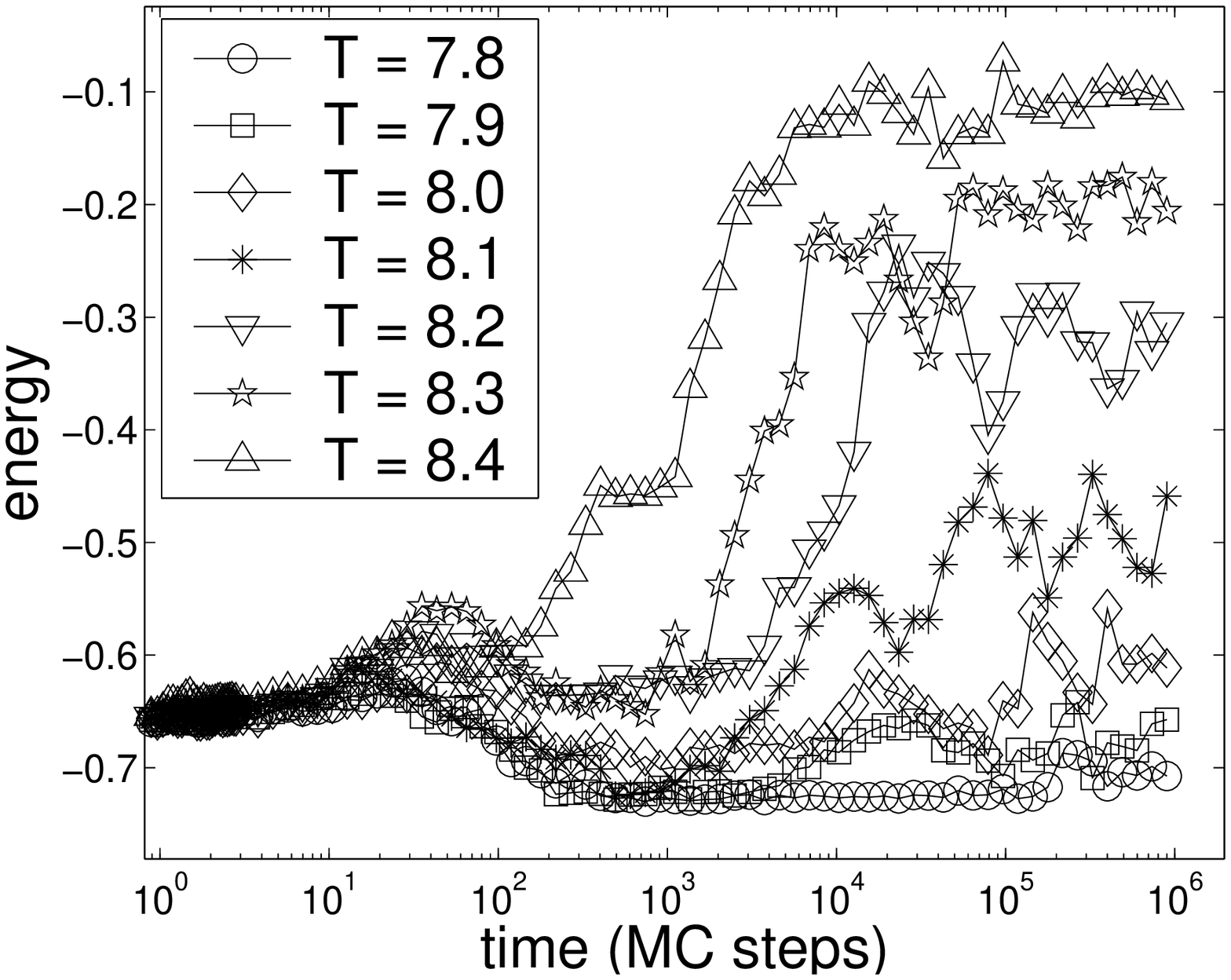}
\includegraphics[width=0.85\columnwidth]{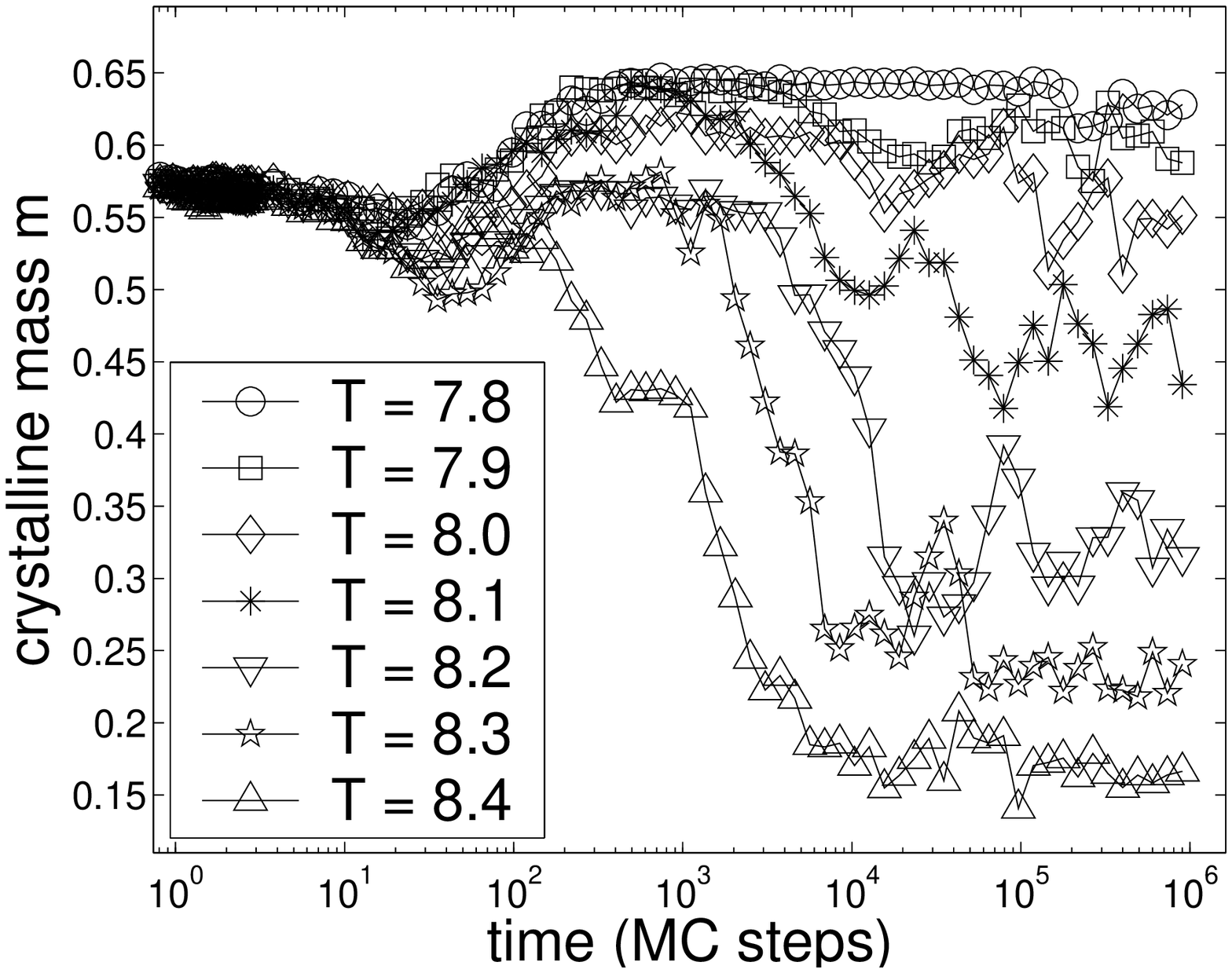}
\caption{
\label{Emvst_T*}
Time evolution of the internal energy and crystalline mass, after the 
system has been prepared in an almost polycrystallized configuration. 
The curves correspond to different quenching temperatures both above and 
below the transition temperature $T^* \simeq 8.1 \pm 0.1$. 
Note that all the temperatures are below the thermodynamic transition 
temperature $T_c = 9.6$, while the system behaves as if it is incapable 
of visiting the favored FMFS configuration. 
}
\end{figure}
In Fig.~\ref{Emvst_T*} (top) we present the results in terms of time evolution 
of the energy. Even though we do not have a sharp distinction between the 
behavior above and below $T^*$, we can clearly identify a transition at 
$T^* \simeq 8.1 \pm 0.1$. When the system is set to a temperature $T > 8.2$, 
it quickly departs from the quasi-polycrystallyzed initial state, while for 
$T < 8.0$ it completes the polycrystallization process, thus lowering its 
energy. 
It is interesting to notice that all the quenching temperatures are below 
the thermodynamic transition temperature $T_c = 9.6$, while the system 
behaves as if it is incapable of visiting the favored FMFS configuration. 

Since the total magnetization of the system remains close to zero 
for all temperatures and time scales that we are able to sample, 
it cannot be used as an order parameter for this transition. A 
more appropriate order parameter is probably the crystalline mass 
$m$, shown in Fig.~\ref{Emvst_T*} (bottom). As proposed by Cavagna 
\textit{et al.}~\cite{Cavagna}, the crystalline mass measures the 
fraction of crystallized spins independently of the size of the 
polycrystals. We first define the elementary crystal unit as 
the four spin cluster composed by one spin and its three 
nearest-neighbors. To avoid double-counting, we choose the central 
spin exclusively in one of the two sublattices of the hexagonal 
lattice. Then, we define the crystal mass density $m \in [0,1]$ as 
the number of these elementary units present in a given 
configuration, normalized by the total number of units $L^2$. 
Since we need to keep the elementary unit small enough to be 
sensitive to small amounts of crystal mass, we have a limited 
power of resolution. In fact, even a random configuration has a 
non-zero average crystalline mass $m_0 = 0.01$, which we consider 
as the effective zero of $m$. The results obtained by measuring 
the time evolution of $m$ are in good agreement with the 
conclusion that $T^* \simeq 8.1 \pm 0.1$. 
\subsubsection{Some considerations on the dynamics of the polycrystal}
The data shown in Fig.~\ref{Emvst_T*} are averages over $32$ 
different histories starting from the same initial configuration. 
The reasons for the large time fluctuations and the lack of a 
sharp distinction between above-$T^*$ and below-$T^*$ behavior, as 
shown instead in the system studied by Cavagna 
\textit{et al.}~\cite{Cavagna}, 
are to be found in the peculiar, rare-event-dominated dynamics of the 
polycrystalline phase. It is worth to analyze this dynamics in detail, 
as it helps understanding also the phase-space isolation of the 
thermodynamic GS, i.e. the FMFS crystal.

With some simple reasoning about the colors and the chirality spins, 
one can see that within a 
single, ferromagnetically ordered domain, all the bonds of the same 
color are aligned in the same direction. Thus, any two-color sequence 
inside the domain follows a straight path from one side to the other 
along one of the three crystalline directions (or crystalline planes) 
of the hexagonal lattice.  This high level of order is responsible for 
the first important difference with respect to usual domain growth: 
there are no small loops across the boundary of a domain (but for 
possible corner loops) and the domain is not capable of small 
rearrangements of its walls. While for example in a normal Ising model 
a domain can expand gradually, in our constrained Ising model a domain 
can only crack from side to side. It is important to notice that these 
cracks will almost always bring the system into an excited state with 
higher energy, the energy difference being proportional to the length 
of the crack. 

\begin{figure*}[!ht]
\begin{center}
\subfigure[\label{corr_PC(a)}
Waiting time $t_w = 20$ MC steps]
{\includegraphics[width=0.75\columnwidth]{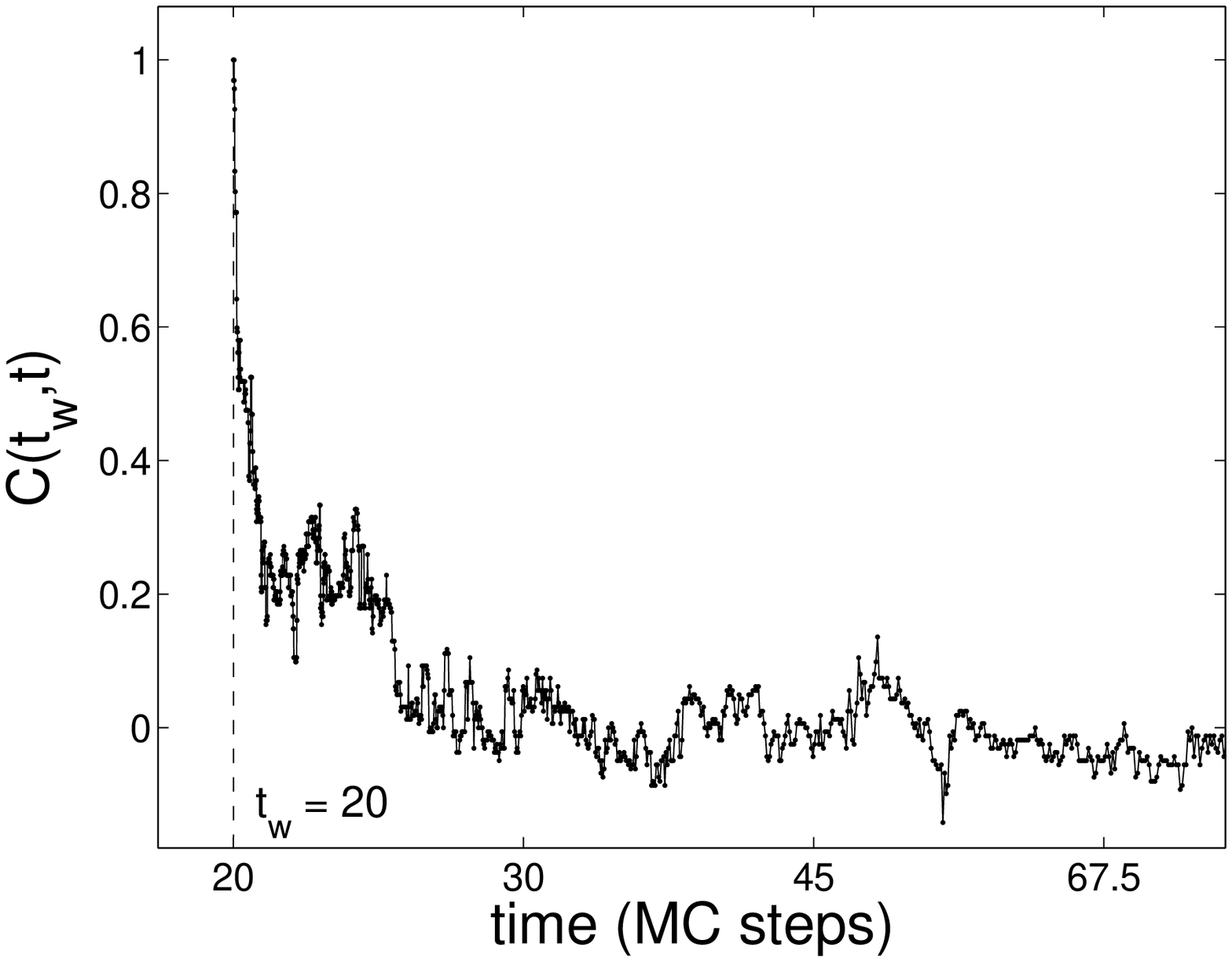}}
\hspace{1 cm}
\subfigure[\label{corr_PC(b)}
Waiting time $t_w = 2\cdot10^2$ MC steps. Note the rescaling of the 
time axis with respect to the previous figure]
{\includegraphics[width=0.75\columnwidth]{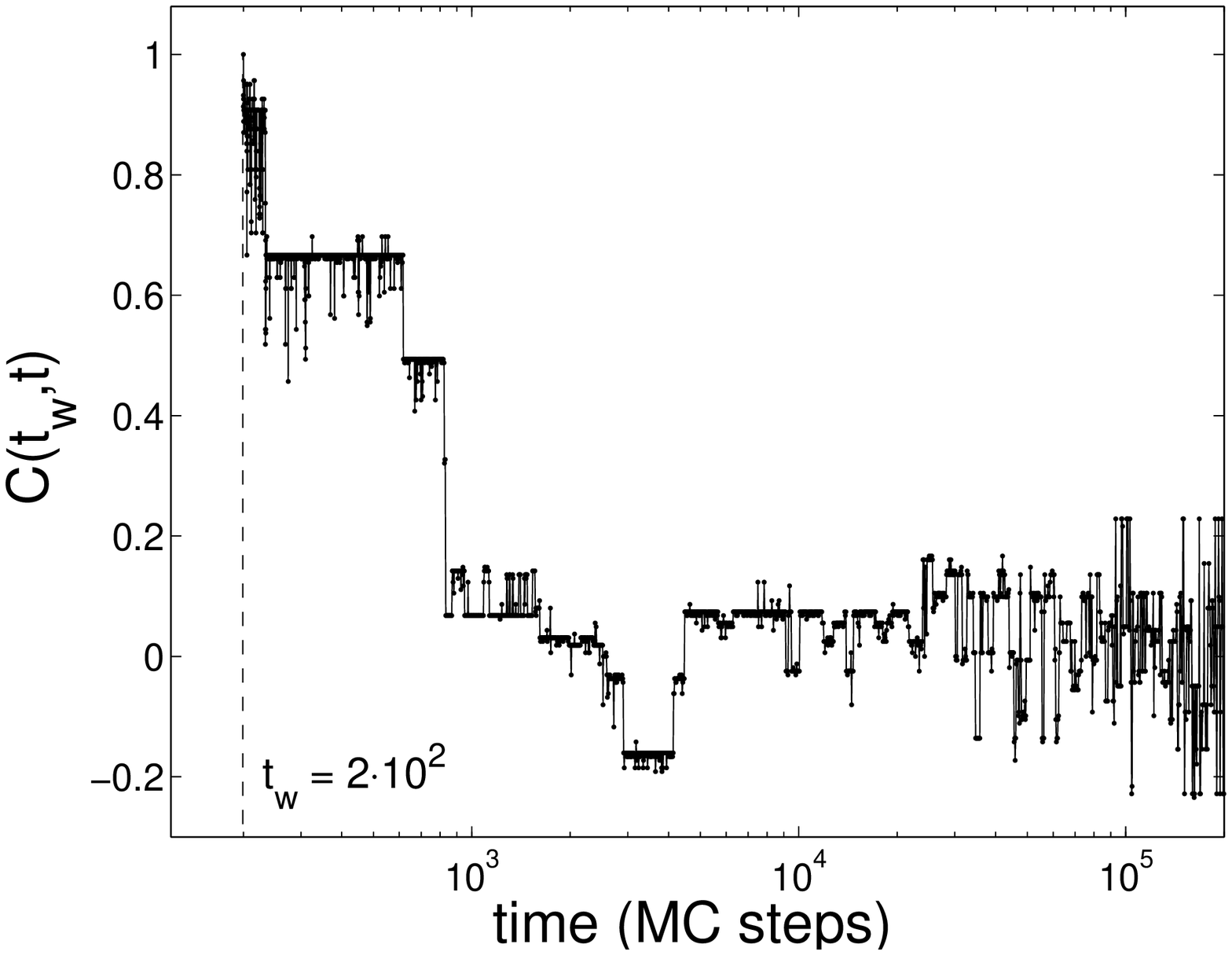}}
\\
\subfigure[\label{corr_PC(c)}
Waiting time $t_w = 2\cdot10^3$ MC steps]
{\includegraphics[width=0.75\columnwidth]{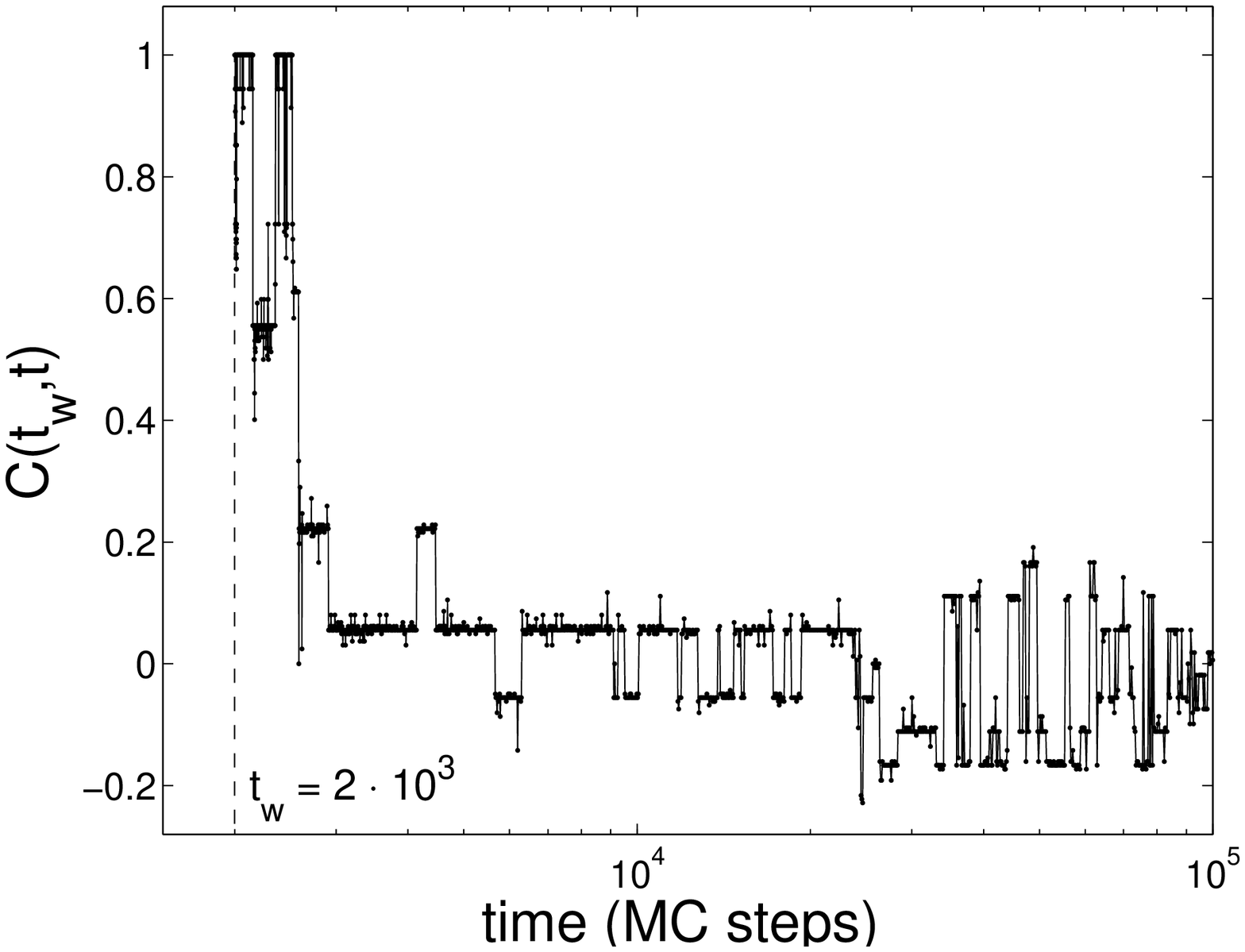}}
\hspace{1 cm}
\subfigure[\label{corr_PC(d)}
Waiting time $t_w = 2\cdot10^4$ MC steps]
{\includegraphics[width=0.75\columnwidth]{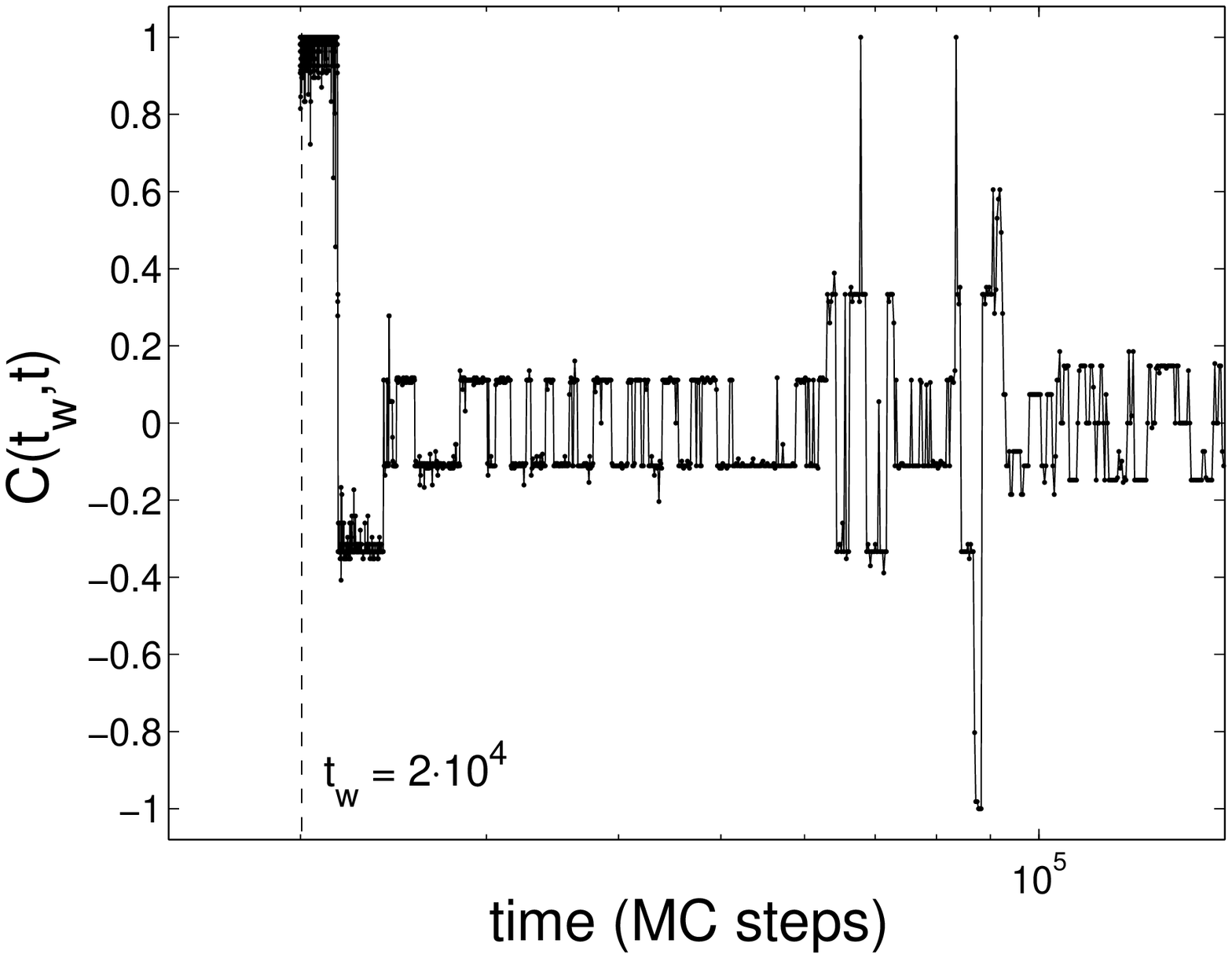}}
\end{center}
\caption{
\label{corr_PC}
Spin-spin autocorrelation function $\mathcal{C}(t_w,t)$ 
for a single MC simulation and four different 
values of $t_w = 20$, $2\cdot10^2$, $2\cdot10^3$ and $2\cdot10^4$ MC 
steps. The temperature is quenched at $t=0$ from $T=\infty$ to $T=6$, 
the same used in Fig.~\ref{evolution}. 
At $t_w = 20$, the system is still in a rapidly-changing liquid phase 
(see Fig.~\ref{evolution(b)}). As the system gets deeper into the 
polycrystalline phase at $t_w = 2\cdot10^3$ or even more at 
$t_w = 2\cdot10^4$ (see Fig.~\ref{evolution(d)}), the behavior of the 
correlation function becomes discontinuous, reflecting a rare event 
dominated dynamics where the system undergoes highly non-local 
rearrangements. 
Notice the $\mathbb{Z}_2$ symmetry of the system (Fig.~\ref{corr_PC(c)}). 
When the dynamics become highly entangled in the polycrystalline phase 
(see Fig.~\ref{evolution(e)}), the number of allowed configurations drops 
dramatically and rearrangements that bring the system from one 
configuration to its mirror image play a significant role in the 
evolution of the system. 
}
\end{figure*}

If we now extend these considerations to the almost complete 
polycrystalline phase that the system is able to achieve below $T^*$ 
(see Fig.~\ref{evolution}), we can see that any loop has to cross 
a few domains before closing on itself. In fact, bending of the 
loops are allowed only at domain boundaries. Therefore, we have a 
second important difference with respect to usual domain growth: 
one domain cannot expand at the expenses of a single other domain; 
rather, the above cracks involve at least six domains (but for the 
case of winding loops), since every domain boundary corresponds to a 
$60^\textrm{o}$ bending in the loop. One can easily convince oneself 
that the closer the system is to the polycrystalline phase, the more 
the dynamics become frozen, requiring entangled, multiple-domain 
cracking in order to move from one configuration to another. 
This behavior can be seen for example by looking at the behavior of 
the spin-spin autocorrelation function (see eqn.(\ref{autocorr})), 
shown in Fig.~\ref{corr_PC}. 
For small values of $t_w$, the system is still in a rapidly-changing 
liquid phase (see Fig.~\ref{evolution(b)}), and the correlation function 
roughly follows the stretched exponential behavior with a very short 
relaxation time discussed in Sec.~\ref{sssec:domliq}. 
As the system gets deeper into the polycrystalline phase 
for $t_w = 2\cdot10^3$ or even more for $t_w = 2\cdot10^4$ (see 
Fig.~\ref{evolution(d)}), the behavior of the correlation function 
shows how the system now evolves mostly via rare 
events that are responsible of extended changes in the system 
configuration. 
Notice the $\mathbb{Z}_2$ symmetry of the system. When the dynamics become 
highly entangled in the polycrystalline phase (see Fig.~\ref{evolution(e)}), 
the number of allowed configurations drops dramatically 
and rearrangements that bring the system from one configuration to 
its mirror image play a significant role in the evolution of the system
(Fig.~\ref{corr_PC(c)}). 
%
%
%
%
%
%

It is important to underline the large energy cost of these updates, which 
scale with the linear size $\xi$ of the domains. Indeed, we can interpret 
this energy difference as the activation energy $E_A(\xi)$ for domain growth. 
Processes where the activation energy depends on $\xi$, or more generally 
where freezing involves a collective behavior dependent on $\xi$ belong to 
classes 3 and 4 for growth kinetics~\cite{Lai}. In the next paragraph, we 
will address this classification in greater detail. 

Even if the system is able to overcome the activation energy 
barrier, the three-coloring constraint plays a new key-role in 
preventing the system from reaching a new configuration. Let us 
consider an excited state after one loop has been updated in the 
polycrystalline phase. The system has then three types of updates 
available: the trivial repair of the crack, with consequent 
lowering of the energy; an independent update, which requires to 
overcome a similar activation energy; and the peculiar loop 
updates that are adjacent to the open crack. Clearly, since a loop 
update corresponds to flipping all the spins along the loop, the 
latter update has a vanishing energy cost because the original 
crack crosses crystalline ordered domains. Thus, the system is 
able, via these adjacent loops, to expand or contract a crack with 
essentially equal probability. Indeed we expect this process to be 
similar in nature to a random walk, with two possible outcomes: 
the crack eventually contracts and closes on itself, or all the 
domains involved in the original crack get essentially flipped, 
with minimal structural change in the original configuration. 
Notice that the last update in this process is of the repair type, 
with the system getting back to a lower energy state. The time to 
complete this process is the lifetime $\tau_d$ of a crack in the 
system, while the formation time of a new crack is determined by 
the activation energy barrier $\tau_f \sim \exp [-\beta E_A(\xi)]$. 
At low temperatures, $\tau_d$ is much shorter than 
$\tau_f$; the system freezes into a specific polycrystalline 
configuration and the dynamics involve only rare events where 
entire domains are flipped simultaneously. At temperatures close 
to $T^*$ instead, $\tau_d$ becomes comparable to $\tau_f$ and 
multiple cracks allow the system to deeply rearrange the domains. 
Notice however that it is still a rare-event dependent dynamics. 
In a typical process of configuration change, the system visits 
highly excited states with complete ``melting'' of extended areas of 
the polycrystal, before freezing again into a new polycrystalline 
configuration . These highly excited intermediary states easily 
become long lived due to the metastability of the liquid phase, 
which has instead very fast dynamics (see Fig.~\ref{liq_relax} and 
the results hereafter). 
\subsection{One-time quantities}
\subsubsection{Energy vs temperature and growth dynamics}
\begin{figure}[!ht]
\begin{center}
\subfigure[\label{EvsT_hist}
Internal energy vs temperature hysteresis for three different values of 
the cooling/heating rate: $r=0.04$, $0.004$ and $0.00004$. 
The hysteretic behavior is typical of a first order phase transition 
and is in good agreement with our measure of $T^*$.]
{\includegraphics[width=0.85\columnwidth]{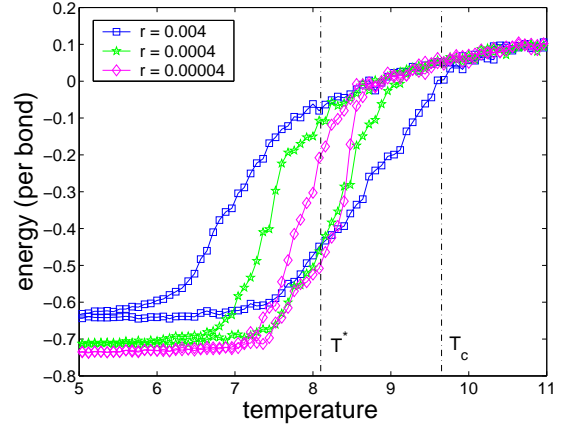}}
\\
\subfigure[\label{EvsT_cool}
Internal energy vs temperature plot for different cooling rates 
$r=0.4$, $0.04$, $0.004$, $0.0004$ and $0.00004$. 
The dashed line is the extrapolated internal energy of the 
liquid phase (eqn.~(\ref{eqn_Elq})). Notice that, 
for $r = 0.4$, the system stays in the liquid phase 
till $T = 0$, since the energy curve remains above the dashed 
line at any temperature~\cite{Cavagna}.]
{\includegraphics[width=0.85\columnwidth]{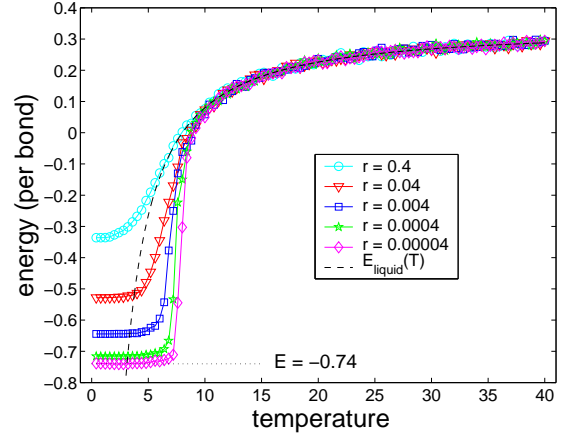}}
\end{center}
\caption{
\label{EvsT_all}
Internal energy vs temperature behavior for our system, in the 
temperature range $T \in (0,40)$. These curves are obtained from 
simulations where the temperature is changed at a constant cooling/heating 
rate. For large temperatures ($T>15$), all the curves overlap and the 
system is at equilibrium in the liquid phase. Notice that there is no sign 
of the thermodynamic transition at $T_c = 9.6$, as the system goes 
smoothly into the supercooled liquid phase. 
}
\end{figure}

In order to get a better insight in the dynamics of the model, we 
study the behavior of the system through temperature hysteresis 
with different cooling/heating rates. We vary the temperature from 
$T=40$, where the liquid phase is stable and equilibrates very easily, 
down to $T=0$ and up again to $T=40$, with a constant rate given 
by $r = \Delta T / \Delta t$, $\Delta t$ being the total time to 
go from $T=40$ to $T=0$. During these simulations we measure all 
the relevant quantities in our system: the internal energy, the 
magnetization, the staggered magnetization, and the crystalline 
mass. Both magnetizations remain close to zero for any temperature 
and cooling/heating rate. The behavior of the internal energy is shown 
in Fig.~\ref{EvsT_all} for some of the cooling/heating rates that we 
consider. The behavior of the crystalline mass is in agreement 
with the internal energy and does not provide any additional 
information. 

The hysteresis observed in the energy curves is typical of first 
order phase transitions. From Fig.~\ref{EvsT_hist} we can see 
that the hysteresis gets narrower for smaller values of $r$, indicating a
transition temperature that is consistent with our previous estimate 
$T^* \simeq 8.1 \pm 0.1$ (that estimate is also confirmed by 
looking at the position of the peaks in the specific heat, measured from 
the energy fluctuations, for different cooling/heating rates). 
Notice the asymmetry of the hysteresis toward the liquid phase, 
particularly evident for large cooling/heating rates, due to the 
metastability of the liquid with respect to the polycrystalline phase. 
For large cooling rates, see for example $r = 0.4$ in 
Fig.~\ref{EvsT_cool}, the energy curves never cross below the 
extrapolated $\mathcal{E}_{\textrm{liquid}}(T)$ curve (dashed line 
in the figure). Thus~\cite{Cavagna}, the system does not 
polycrystallize and it remains in a supercooled liquid phase with 
respect to the polycrystalline phase until $T=0$ (recall that the 
liquid is already supercooled with respect to the FMFS phase for 
$T<9.6$). This is confirmed also by the absence of a peak in the 
specific heat curves. As the temperature is lowered to zero, the curves 
reach a final value of the energy that decreases monotonically with 
smaller cooling rates. But for very large values of $r$ (larger than 
$0.4$), this final value of the energy is reached already at a 
finite temperature and the curves show a plateau typical of frozen 
or very slow dynamics. While we expect this behavior when the 
system enters the polycrystalline phase, we can notice that this 
plateau is also present for curves where the system remains in the 
supercooled liquid phase (e.g. see the curve for $r=0.4$ in 
Fig.~\ref{EvsT_cool}). A detailed analysis of this behavior is 
beyond the scope of the present paper and will be addressed in the 
future. 
 
\begin{figure}[!ht]
\centering
\includegraphics[width=0.85\columnwidth]{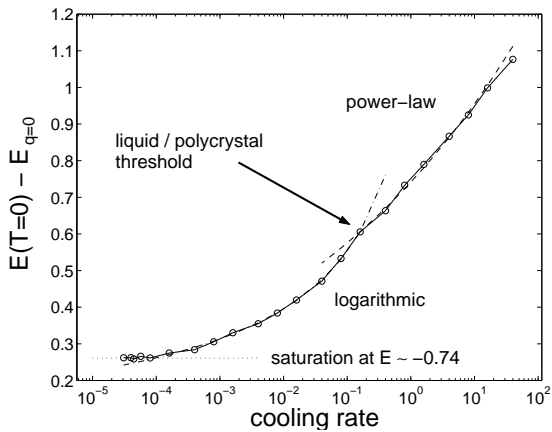}
\caption{
\label{EvsT_plateau}
Semi-logarithmic plot of the plateau value of the internal energy with 
respect to the GS energy of the perfect crystal 
($\mathcal{E}_{\textrm{FMFS}} = -1$), versus the cooling rate 
$r = \Delta T / \Delta t$. Three distinct behaviors can be identified: 
a power law behavior $\mathcal{E} \sim r^{0.11}$ for $r > 0.2$, when the 
system remains in the liquid phase; a logarithmic behavior 
$\mathcal{E}^{-1} \sim \ln (1/r^{0.85})$ for 
$8\cdot10^{-5} \leq r \leq 0.2$; and a saturation plateau at 
$\mathcal{E}^{\textrm{P-xtal}}(T=0) \sim -0.74$ for $r<8\cdot10^{-5}$. 
}
\end{figure}

The dependence of the $T=0$ value of the energy on the cooling 
rate reflects the type of domain growth in the system. In 
particular, when the system enters the polycrystalline phase where 
domain boundaries are one-dimensional, the energy difference 
$\mathcal{E}(T=0)-\mathcal{E}_{\textrm{FMFS}} = \mathcal{E}(T=0)+1$ is 
proportional to the inverse of the linear size of the 
domains~\cite{Cavagna}. In Fig.~\ref{EvsT_plateau} we show the 
behavior of $\mathcal{E}(T=0)-\mathcal{E}_{\textrm{FMFS}}$ as a function of 
$r$. 

As long as the system remains in the liquid phase, i.e. the energy curves 
never cross below the extrapolated $\mathcal{E}_{\textrm{liquid}}(T)$ curve, 
the energy follows a power law dependence on $r$: 
$\mathcal{E}-\mathcal{E}_{\textrm{FMFS}} \sim r^{0.11}$. 
This is typical of class 1 growth kinetics, where freezing originates from 
local defects with activation energies independent of the domain size 
$\xi$~\cite{Cavagna,Lai}. 

As we lower the cooling rate, we reach a threshold where the energy curves 
start crossing the extrapolated $\mathcal{E}_{\textrm{liquid}}(T)$ curve 
and the system polycrystallizes. This threshold happens at 
$r_{\textrm{th}} \simeq 0.2$ and $\mathcal{E}_{\textrm{th}} \simeq -0.39$. 
Below this threshold, the behavior of the energy changes abruptly into a 
logarithmic form: 
\beq
\mathcal{E}(T=0)-\mathcal{E}_{\textrm{FMFS}} =
  \frac{1}{1 + A \left[ \ln(\frac{1}{r \cdot \tau_1}) \right]^m}.
\eeq
From a fit of the results we obtain $m \simeq 0.85$, even though 
our numerical data do not have enough accuracy to exclude the case $m=1$. 
If our measurement of $m \neq 1$ is confirmed, it implies that the behavior 
of our system for $r \in [8\cdot10^{-5},0.2]$ belongs to class 4 growth 
kinetics~\cite{Lai}. Both class 3 (corresponding to the case of $m=1$) and 
class 4 kinetics are typical of processes that involve a $\xi$-dependent 
collective behavior in the frozen phase. As discussed above, we indeed 
expected the system to show this logarithmic behavior. 

Eventually, for $r < 8\cdot10^{-5}$ the energy saturates to a 
limiting value $\mathcal{E}^{\textrm{P-xtal}}(T=0) \sim -0.74$, in 
agreement with the entropic argument we provided before. The 
system behaves as if a whole region of phase space around the FMFS 
configuration is dynamically inaccessible due to a very large free 
energy barrier. 

To further confirm this peculiar free energy landscape, we use 
again the CMFM described in Sec.~\ref{CFMsec}. From the numerical 
results, we assume as a first-order approximation that the 
dynamically excluded configurations correspond to system energies smaller 
than the limiting value $\mathcal{E}^{\textrm{P-xtal}}(T=0)\sim -0.74$. 
We then impose appropriate constraints on the variational parameter such 
that the only allowed energies in the CMFM are larger than 
$\mathcal{E}^{\textrm{P-xtal}}(T=0)$. 
Under these constraints, the method predicts a first order phase 
transition at $T^* \simeq 8.36$, in good agreement with 
the numerical value $T^* \simeq 8.1 \pm 0.1$, considering the 
approximations underlying this CMFM result. 
%
%
%
\subsubsection{
\label{sssec:domliq}
Domain nucleation vs liquid relaxation}
Here we study the equilibration time of the liquid phase in 
comparison to the nucleation time for the polycrystalline phase. 

\begin{figure}[!ht]
\centering
\includegraphics[width=0.85\columnwidth]{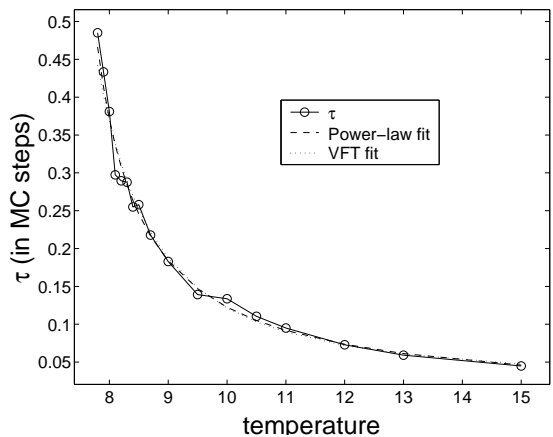}
\caption{
\label{liq_relax}
Liquid phase relaxation time $\tau$ as a function of temperature, as measured 
from the stretched exponential fit of the autocorrelation function at 
equilibrium. The dashed line corresponds to a power law fit while the dotted 
line corresponds to a Vogel-Fulcher-Tamman fit. Notice that there is no 
dynamic signature of the transition at temperature $T^*$. 
}
\end{figure}

We measure the connected piece of the two-times autocorrelation function: 
\beq
\mathcal{C}(t_w,t) = \frac{1}{2 L^2} \sum_i
   \langle \sigma_i(t_w)\sigma_i(t) \rangle,
\label{autocorr}
\eeq
where $\langle \dots \rangle$ indicates the average over initial 
configurations of different MC simulations. 
Notice that $\sum_i \sigma_i(t) \simeq 0$ for all values of $t$ within our 
simulation time scale, thus the disconnected piece of the autocorrelation 
function vanishes. 
Since we are interested in the relaxation time of the liquid 
phase at equilibrium, we quench the system from infinite temperature down 
to the target temperature T and we wait for it to equilibrate. The 
correlation function becomes time-translation invariant and depends only 
on the time difference $t-t_w$. At equilibrium, we adequately fit 
$\mathcal{C}(t-t_w)$ with a stretched exponential, which is the expected 
equilibrium behavior in supercooled liquids~\cite{Cavagna}: 
\beq
\mathcal{C}(t) = \exp{[-(t/\tau)^\beta]}.
\eeq
From the fit we obtain the relaxation time $\tau$ as a function of the 
quenching temperature, as shown in Fig.~\ref{liq_relax}. 
We can extend the measurement of $\tau$ below $T^*$ because of the 
metastability of the liquid phase. The system is able to 
equilibrate as a supercooled liquid well before the polycrystal 
transition takes place, at least for temperatures close enough to 
$T^*$. Notice that there is no dynamic signature of the 
polycrystal transition at $T^*$ in the liquid relaxation time. The 
Kohlrausch exponent $\beta$ of the stretched exponential fit 
decreases with temperature, as for realistic models of liquids. 
In Fig.~\ref{liq_relax} we show the fit of the $\tau$ 
data both with a power law: 
\beq
\tau = \frac{A}{(T-T^{\textrm{lq}}_c)^\gamma},
       \;\;\;T^{\textrm{lq}}_c=7.0\;\;\gamma=1.0,
\eeq
and with a Vogel-Fulcher-Tamman (VFT) form: 
\beq
\tau=\tau_0\exp{\left( \frac{\Delta}{T-T_0} \right)},
     \;\;\;T_0=4.4\;\;\Delta=11.1.
\eeq
The results of these fits have to be considered with extreme care. 
Because of the accelerated non-local dynamics and because of the onset 
of polycrystal nucleation, the temperature range where we are able to 
measure the relaxation time of the liquid phase allows for $\tau$ to vary 
only over a narrow interval, from $0.05$ to $0.5$ MC steps. 
As a consequence, the values obtained for the fitting parameters lack in 
accuracy, since the fit spans a single decade of data. 
Moreover, a VFT behavior typically involves the large $\tau$ limit of the 
$\tau(T)$ curve, which is not accessible in the present system due to the 
rapid nucleation of the polycrystal. 
Indeed, our numerical data are the tail of a possible VFT behavior, and they 
suggest that a VFT behavior may be observed in the liquid 
phase of this system if the polycrystallization process were to be avoided. 

Since the correlation function decays to zero in approximately 
$20\,\tau$, we can take this value as the equilibration time 
for the liquid phase at a given temperature~\cite{Cavagna}: 
$\tau_{\textrm{eq}}(T) = 20\,\tau(T)$. 

\begin{figure}[!ht]
\centering
\includegraphics[width=0.85\columnwidth]{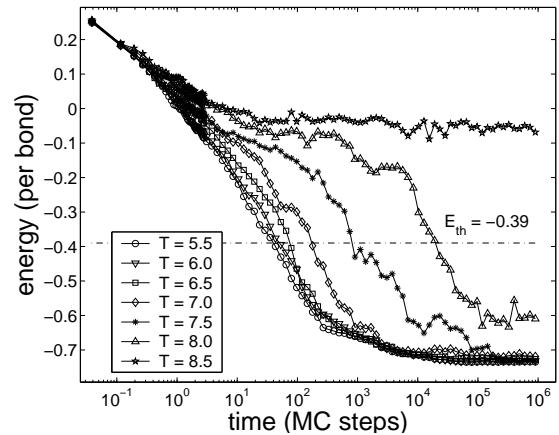}
\caption{
\label{tau_nucl}
Time evolution of the energy of the system, after a quench from infinite 
temperature down to a target temperature $T = 5.5, 6.0, 6.5, 7.0, 7.5, 8.0$ 
and $8.5$. The horizontal line corresponds to the energy threshold for 
polycrystallization $\mathcal{E}_{\textrm{th}} \simeq -0.39$, as identified 
above. 
}
\end{figure}

Measuring the nucleation time of the polycrystalline phase in this system 
is instead more complicated. Due to the frozen nature of the polycrystalline 
phase, we cannot compute its free energy as a function of temperature as 
we did for the liquid phase (see Sec.~\ref{sssec:tt_dyn}). Thus, methods 
such as the one in Cavagna \textit{et al.}~\cite{Cavagna} are not 
applicable. More naively, we have to estimate $\tau_{\textrm{nucl}}$ 
directly observing the time evolution of the system. In Fig.~\ref{tau_nucl} 
we plot the energy dependence on time for quenches of the system from 
infinite temperature to the target temperature $T$. 
As we discussed above, the system polycrystallizes when the energy 
falls below a threshold value $E_{\textrm{th}} \simeq -0.39$. Here we use 
this value in order to identify the onset of the 
polycrystallization process in the energy curves in 
Fig.~\ref{tau_nucl}. The time when the system starts developing a 
polycrystalline phase is indeed the nucleation time $\tau_{\textrm{nucl}}$ we 
are interested in. We can see that $\tau_{\textrm{nucl}} \simeq 800$ at 
$T = 7.5$ while it drops to $\tau_{\textrm{nucl}} \simeq 170$ at $T = 7.0$. 

Comparing these results with the ones of Fig.~\ref{liq_relax}, provided we 
perform the rescaling $\tau_{\textrm{eq}} = 20\,\tau$, we can see that 
the crossover $\tau_{\textrm{eq}} = \tau_{\textrm{nucl}}$ will happen 
at a temperature $T_{sp}$ close to $T^{\textrm{lq}}_c$, where the liquid 
relaxation time shows a rapid growth. We can reasonably locate this 
crossover in the temperature range $7.0 < T_{sp} < 7.5$. 
This temperature is the spinodal 
temperature corresponding to the metastability limit of the liquid, when 
the liquid equilibration timescales become of the same order of the 
nucleation timescales and the liquid phase becomes unstable. 
The system reaches this limit in a time $t_{sp}$ of 
the order of a few hundred MC steps. 

%
%
\section{\label{sec:conclusions}Conclusions}
In this paper we have studied the very interesting properties of a
model for describing the behavior, both static and dynamic, of
different arrays of superconducting devices. Among the examples
discussed, the main candidate to see such a rich phenomenology is a
Josephson junction array of triangular grains of superconductors with
$p_x \pm i p_y$ order parameter. In the limit of very strong Josephson
couplings, the system is equivalent to Baxter's three color model in
the hexagonal lattice. This model can in turn be represented by an
Ising model with a constraint on the total magnetization for each
hexagonal plaquette, $\sigma^{\mbox{\small\hexagon}}_P = \pm 6,0$. In
this paper we have presented a proof of this mapping based on the
condition of the singlevaluedness of a superconducting order
parameter. The Ising degrees of freedom correspond, in the Josephson
arrays with $p$-wave islands, to the chirality of the $p_x \pm i p_y$
order parameter.

Within the constrained $\sigma^{\mbox{\small\hexagon}}_P = \pm 6,0$
space, the system is critical at infinite temperature but orders at
any finite temperature if antiferromagnetic interactions between the
Ising spins are present. For ferromagnetic interactions, it remains
critical until a very particular first order phase transition takes
place, where the system orders completely. This behavior is due to the
peculiar nature of the ordered state, which is isolated in phase space
from any of its excitations by an energy of order the system
size. 


For a finite Josephson coupling strength, defects are present in the
system, and there are violations of the color and, consequently,
$\sigma^{\mbox{\small\hexagon}}_P = \pm 6,0$ plaquette constraint. A
particularly interesting kind of defect is a fractional vortex
pair. Within the context of the Josephson array of $p_x \pm i p_y$
superconducting islands, not only there is a large energetic cost to
create these excitations, but they are also confined at low
temperatures by logarithmic interactions. The other kind of
interesting excitation is formed by flipping the spins along open
segments of closed two-color loops. While there is also an energetic
cost to create them, these defects can circulate on the lattice
without further energetic cost, in contrast with the fractional
vortices.  Moreover, a new defect-free color configuration is obtained
through the process of creation of a string of spin flip excitations,
the propagation of the defect along the two-color loop, and the
recombination of the ends of the string after closing the loop. This
mechanism is precisely the microscopic origin of the Monte Carlo
dynamics that we implement in this paper.

Because of the constraint, the dynamics of the system is very
peculiar. While the existence of a super-cooled liquid phase is
typical of first order transitions, for our constrained system we find 
a whole temperature range in which such super-cooled liquid is stable 
for extremely long time scales. Indeed, at all the time scales studied 
in this paper, the fully order phase can not be reached and the system 
orders into a polycrystalline phase in which the global $\mathbb{Z}_2$ 
symmetry is unbroken. The transition from the liquid state to the 
polycrystal takes place at a critical temperature $T^*$, smaller than 
the static (avoided) critical temperature. This dynamical temperature
$T^*$ has been obtained both by studying the time evolution of the
system after preparing it in a polycrystalline configuration, and by
quenching the system from the liquid phase. The values obtained for
$T^*$ are in agreement with the naive estimate that we are able to
obtain from the CMFM technique.  The numerical analysis of the
nucleation time and the liquid phase relaxation time allows us to give
an estimate of the spinodal temperature of the liquid. 

The rich phenomenology of the dynamics of this system is also 
reflected in the dependence of the difference between the final 
internal energy reached by the system and that of the fully ordered 
state on the cooling rate. While for very fast cooling rates this 
dependence shows a typical power law scaling, the nucleation of the 
polycrystalline phase produces a logarithmic behavior until a total 
arrest in the domain growth is reached, meaning probably another 
logarithmic growth but with a much longer time scale. 
The origin of this scenario is the fact that the energy 
barriers through which the system has to pass to reach 
states with larger clusters grow with the size of the clusters. This 
places our system as one of the rare cases without randomness in which 
the dynamics is of class 3 or 4 in the classification of Lai 
\textit{et al.}~\cite{Lai}. 

An important open problem concerns the possible mechanism to get out
of the polycrystalline state. Proliferation of other (confined) type
defects, such as fractional vortices, is a possible mechanism to help
overcome the totally arrested dynamics in the polycrystalline
phase. In this case, the large time scale dynamics could be governed
by the energetic cost of making a rather rare event dominated
proliferation and circulation of such (confined) defects. It is
noteworthy that, in the polycrystalline phase, not only the fractional
vortices are confined (logarithmically, with a prefactor of order
$U$), but also the excitations that we argued are responsible for the
microscopic dynamics, the open segments of closed two-color loops. The
confinement of the two-color segments is proportional to the string
length (linear) inside any ferromagnetically aligned domain, with a
prefactor of order $J$. The example that we studied in this paper
suggests an interesting scenario where defect confinement at the
microscopic level is responsible for the slow dynamics and
out-of-equilibrium behavior of a macroscopic system.

\begin{acknowledgments}
We would like to thank H. Castillo and M. Kennett for enlightening 
discussions. We are particularly thankful to B. Chakraborty, D. Das, 
and J. Kondev for several stimulating discussions that greatly 
motivated our interest in this problem, and J. Moore for several 
discussions and correspondence on the issue of superconducting 
realizations of the model. This work is supported in part by the NSF 
grant DMR-0305482 (C.~C. and C.~C.). P.~P. would like to thank the 
warm hospitality of the Boston University Physics Department, where 
this work was carried out. 
\end{acknowledgments}
\appendix
%
%
\section{\label{AppB}Microscopic origin of the $U$ and $J$ terms}
In this appendix we estimate the relative values of $U$ and $J$ in 
terms of some microscopics for the tunneling through a Josephson 
barrier. Consider two neighboring triangles as in 
Figure~\ref{triangle} sharing a common edge labelled by $a$. The 
microscopic tunneling Hamiltonian from a triangle labelled $1$ to a 
neighboring triangle labelled $2$ can be written as: 
\beq
H = - \sum_{\vec{k},\vec{q}} 
      t_{\vec{k},\vec{q}} ~ 
      c^{1 \dagger}_{\vec{k}}c^2_{\vec{q}} \; + \; \textrm{h.c.} 
\eeq
Using second order perturbation theory, we can estimate from this expression  
the Josephson coupling between the two superconductors in a standard way.  
The result is: 
\beq
E_J \sim  - \sum_{\vec{k},\vec{q}} 
  \frac{\left| t_{\vec{k},\vec{q}} \right|^2}{E_\Delta} 
\langle c^{1 \dagger}_{-\vec{k}} \, c^{1 \dagger}_{\vec{k}} \rangle  
\langle c^{2}_{\vec{q}} \, c^{2}_{-\vec{q}} \rangle
\; ,
\label{eq:b2}
\eeq
where we used $t_{\vec{k},\vec{q}}=t^*_{-\vec{k},-\vec{q}}$ and 
$E_\Delta$ is the superconducting energy gap. 

It is useful to define the angles $\phi_{\hat k}$ and $\phi_{\hat q}$ 
as those formed by the vectors $\vec k, \vec q$ and the reference unit 
vector $\hat e_{1,0}$. Notice that the $a$-th unit vector $\hat e_{i,a}$ 
is normal to the side labelled by $a$ (see Fig.~\ref{triangle} 
for the definition of these unit vectors). 

The order parameters can be written as: 
\bea
\langle c^{1 \dagger}_{-\vec{k}} \, 
        c^{1 \dagger}_{\vec{k}} \rangle &=&
\left(\Delta^1_{\vec{k}} 
\right)^* = \Delta\;
e^{ -i \left( \theta_1 + \sigma_1 \phi_{\vec{k}} \right)} 
\label{eq:b3}
\\
\langle c^{2}_{\vec{q}} \, 
        c^{2}_{-\vec{q}} \rangle &=& 
\;\;\Delta^2_{\vec{q}}\;\;\;\;\, = \Delta\;
e^{ i \left( \theta_2 + \sigma_2 \phi_{\vec{q}} \right)},
\label{eq:b4}
\eea
where $\Delta$ is the order parameter magnitude, $\theta_{1,2}$ are 
the overall phases of grains $1,2$, and $\sigma_{1,2}$ are the 
chiralities of the $p\pm i p$ order parameter in each grain. 

As we show below, the constants $U$ and $J$ strongly depend on the 
behavior of $t_{\vec{k},\vec{q}} $, which is in general very difficult 
to obtain from first principles. For a flat interface, the component 
of momentum parallel to the junction is conserved, i.e. 
$k_{\parallel}=q_{\parallel}$. If the momenta involved are close to 
the Fermi momentum (and assuming for simplicity a spherically symmetric 
Fermi surface), then one has (approximately) that 
$k_{\parallel}^2+k_{\perp}^2\approx k_F^2\approx q_{\parallel}^2+q_{\perp}^2$; 
hence, $k_{\perp}\approx q_{\perp}$ or $k_{\perp}\approx - q_{\perp}$, 
corresponding to forward and backward scattering in the normal direction 
to the barrier, respectively. 

There should be a strong suppression of tunneling when the vectors 
$\vec k$, $\vec q$ are not normal to the interface. The reason is that 
the smaller the perpendicular component, the more exponentially 
suppressed is the tunneling amplitude (for example, consider a WKB 
approximation: the smaller $k_{\perp}$, $q_{\perp}$, the deeper under 
the barrier). If $\delta\varphi$ is a small angle that measures 
deviations from normal incidence and 
$\phi_{\hat e_{1,a}}=\frac{2\pi}{3} a$, one can show that the main 
contribution to the Josephson tunneling Hamiltonian comes from choosing 
any of the following four combinations: 
\begin{equation}
\phi_{\hat k}=\frac{2\pi}{3} a +\delta\varphi
\qquad {\rm or} \qquad
\phi_{\hat k}=\frac{2\pi}{3} a +\delta\varphi +\pi
\end{equation}
and  
\begin{equation}
\phi_{\hat q}=\phi_{\hat k}
\qquad {\rm or} \qquad 
\phi_{\hat q}=\phi_{\hat k}+\pi -2\delta\varphi 
\; ,
\end{equation}
where the last choice corresponds to forward or backward scattering 
respectively. 

The Josephson coupling can be written in terms of these choices as: 
\begin{eqnarray}
E_J &\sim& \;  - \frac{\Delta^2}{E_\Delta}
\int d\,{\delta\varphi} \cdot
\nonumber\\
&&\hspace{-1 cm} \;
\Big[ |t_F(\delta\varphi)|^2
\;\cos(\theta_{1,a}-\theta_{2,a}+(\sigma_1-\sigma_2)\,\delta\varphi)
\nonumber\\
&&\hspace{-1 cm}
-|t_B(\delta\varphi)|^2
\;\cos(\theta_{1,a}-\theta_{2,a}+(\sigma_1+\sigma_2)\,\delta\varphi)
\Big] ,
\end{eqnarray}
where $t_F(\delta\varphi)$, $t_B(\delta\varphi)$ are forward and backward 
small angle scattering amplitudes (also, recall the definition 
$\theta_{i,a}=\theta_i+\frac{2\pi}{3} a\,\sigma_i$ from section 
\ref{sec:model}). 

Expanding around small $\delta\varphi$ before carrying out the angular 
integral, one obtains: 
\begin{equation}
E_J \sim  - [U\;+J\;\sigma_1\sigma_2]\; \cos(\theta_{1,a}-\theta_{2,a})
\; ,
\end{equation}
where 
\begin{equation}
U=\frac{\Delta^2}{E_\Delta}\int d\,{\delta\varphi} 
\left[|t_F(\delta\varphi)|^2-|t_B(\delta\varphi)|^2\right]\,
(1-\delta\varphi)^2
\end{equation}
and 
\begin{equation}
J=\frac{\Delta^2}{E_\Delta}\int d\,{\delta\varphi} 
\left[|t_F(\delta\varphi)|^2+|t_B(\delta\varphi)|^2\right]\,
\delta\varphi^2
\; .
\end{equation}

As we discussed above, the barrier is more transparent for close to 
normal incidence, and can be engineered so that $\delta\varphi$ must 
remain small, and thus the ratio $J/U$ as obtained above can be made 
controllably small. The precise condition for having $J\ll U$ depends 
on the details of $t_{F,B}(\delta\varphi)$. As a simple example, for 
tunneling through a square barrier in ordinary quantum mechanics, the 
ratio $J/U$ will depend on the height of the barrier $V$ and on $k_F 
a$, where $a$ is the length of the barrier. The larger $k_F a$, the 
smaller $J/U$. This model may not capture in full detail the 
underlying physics of the Josephson coupling problem~\cite{Mahan}; 
nevertheless, simple as it is, it shows how the structure of the 
barrier can be used to tune the ratio $J/U$. 

If $J\ll U$, then in the temperature regime $J\ll T\ll U$ the system 
is effectively constrained to the three-color manifold of states: 
$\theta_{1,a}-\theta_{2,a}=0\; ({\rm mod} 2\pi)$. In this case, the 
effective Hamiltonian for the coupling between triangles $1,2$ is 
simply: 
\begin{equation}
H_{1,2}=-J\sigma_1\sigma_2
\; ,
\end{equation}
with $J>0$ (ferromagnetic coupling). 
%
%

%
%

\begin{thebibliography}{99}
\bibitem{Plaquette_note} 
Here by plaquette we mean any closed polygon of spins. The case of more 
complex spin plaquettes (e.g. with internal spins) requires some non-trivial 
adjustments in the CMFM.
\end{thebibliography}
\end{document}